\documentclass[11pt]{article}

\usepackage[normalem]{ulem}

\usepackage{xcolor}
\usepackage[english]{babel}
\usepackage[T1]{fontenc}
\usepackage[utf8]{inputenc}
\usepackage{authblk}
\usepackage{mathtools}
\usepackage{slashed}
\usepackage{amsmath,amssymb}
\usepackage{amsfonts}
\usepackage{enumitem}
\usepackage{graphicx,color,xcolor}
\usepackage{wrapfig}
\usepackage{cite}
\usepackage{subcaption}
\usepackage{hyperref}
\hypersetup{
	colorlinks=true,
	linkcolor=blue,
	filecolor=magenta,
	urlcolor=cyan,
}
\usepackage{cleveref}

\numberwithin{equation}{section}

\definecolor{refcol}{rgb}{0.9,0.1,0.1}
\hypersetup{colorlinks=true,linkcolor=blue,citecolor=refcol,urlcolor=cyan,linktocpage}

\graphicspath{{Plots/}}

\newcommand{\cA}{\mathcal{A}}

\begin{document}
	\begin{titlepage}
		\thispagestyle{empty}
		
		\title{
			{\Huge\bf  Generalized Freudenthal duality for rotating extremal black holes}
		}
		
		\vfill
		
		\author{
			{\bf Arghya Chattopadhyay$^a$}\thanks{{\tt arghya.chattopadhyay@umons.ac.be}},
			{\bf Taniya Mandal$^b$}\thanks{{\tt
taniyam@iitk.ac.in}},
			{\bf Alessio Marrani$^{c}$}\thanks{{\tt alessio.marrani@um.es}}
			\smallskip\hfill\\      	
			\small{
				
				${}^a${\it Service de Physique de l'Univers, Champs et Gravitation}\\
				{\it Université de Mons, 20 Place du Parc, 7000 Mons, Belgium}
				\hfill\\
				\smallskip\hfill\\
                ${}^b${\it Department of Physics, Indian Institute of Technology Kanpur, }\\
                {\it  Kanpur 208016, India}\\
				\hfill
				\smallskip\hfill\\
				$^c${\it Instituto de F\'{\i}sica Teorica, Dep.to de F\'{\i}sica}\\
				{\it Universidad de Murcia, Campus de Espinardo, E-30100, Spain}\\
			}
		}

		\vfill
		
		\date{
			\begin{quote}
				\centerline{{\bf Abstract}}
				{\small
					Freudenthal duality (FD) is a non-linear symmetry of the Bekenstein-Hawking entropy of extremal dyonic black holes (BHs) in Maxwell-Einstein-scalar
theories in four space-time dimensions realized as an anti-involutive map
in the symplectic space of electric-magnetic BH charges. In this paper, we
generalize FD to the class of rotating (stationary) extremal BHs, both in
the under- and over-rotating regime, defining a (generalized) rotating FD
(generally, non-anti-involutive) map (RFD), which also acts on the BH
angular momentum. We prove that the RFD map is unique, and we compute the
explicit expression of its non-linear action on the angular momentum itself.
Interestingly, in the non-rotating limit, RFD bifurcates into the usual,
non-rotating FD branch and into a spurious branch, named \textquotedblleft
golden\textquotedblright\ branch, mapping a non-rotating (static) extremal
BH to an under-rotating (stationary) extremal BH, in which the ratio between
the angular momentum and the non-rotating entropy is the square root of the
golden ratio. Finally, we investigate the possibility of inducing transitions
between the under- and over- rotating regimes by means of RFD, obtaining a
\textit{no-go} result.
				}
			\end{quote}
		}
		

	
\end{titlepage}
\thispagestyle{empty}\maketitle\vfill \eject

\baselineskip=18pt

\newpage

\tableofcontents

\newpage

\section{Introduction}

Extremal black holes (BHs) are objects of great interest to the string
theorists, though they concern about a pretty ideal situation, as they lack
temperature $T$. On the other hand, astrophysical BHs are never exactly
extremal, but, for instance, the BH GRS1915+105 observed through X-ray and a radio telescope is likely within $1\%$ of the extremal value of its angular
momentum \cite{McClintock:2006xd}; in other words, this BH is \textit{%
near-extremal}, i.e. not far from saturating the extremality bound.

For asymptotically flat and static solutions, \textit{Freudenthal duality}
(FD) is an intrinsically non-linear and anti-involutive map between the BH
dyonic charges, which, unexpectedly, keeps the Bekenstein-Hawking BH entropy
invariant \cite{Borsten:2009zy,Ferrara:2011gv,Borsten:2012pd}.
Interestingly, under FD, the attractor configurations of the scalar fields at
the electric-magnetic (e.m.) dyonic BH event horizon also remain invariant
\cite{Ferrara:2011gv}. A suitable extension of FD has been formulated for
Maxwell-Einstein theories in the presence of gaugings of the isometries of
the vector multiplets' or hypermultiplets' scalar manifolds (\textit{at least%
} as far as Abelian gaugings are concerned \cite{Klemm:2017xxk}).
Furthermore, in recent years FD has turned out to characterize a number of
directions of investigation within the Maxwell-Einstein (super)gravity
theories coupled to non-linear sigma models of scalar fields \cite%
{Marrani:2012uu,Galli:2012ji,Fernandez-Melgarejo:2013ksa,Mandal:2017ioi,Borsten:2018djw,Borsten:2019xas}%
. A crucial point to stress is that FD is inherently different from the
electro-magnetic duality (\textit{aka} $U$-duality in string theory\footnote{%
Here $U$-duality is referred to as the \textquotedblleft
continuous\textquotedblright\ symmetries of \cite%
{Cremmer:1978ds,Cremmer:1979up}. Their discrete versions are the U-duality
of the non-perturbative string theory symmetries introduced by Hull and
Townsend in \cite{Hull:1994ys}.}), since the later act \textit{linearly} on
dyonic BH charges, contrary to the former \cite%
{Marrani:2015wra,Marrani:2017ihg,Marrani:2019zsn}.

Recently, in \cite{Chattopadhyay:2021vog} and \cite{Chattopadhyay:2022ycb},
FD has been studied and further extended to the class of \textit{%
near-extremal} (non-rotating) BHs. As mentioned above, for extremal,
(asymptotically flat) non-rotating BHs, FD is an anti-involutive non-linear
map acting on the e.m. BH charges, and it is a symmetry of the
Bekenstein-Hawking BH entropy, which is a homogeneous function of degree two
in the charges themselves. Since the $T$-dependent entropy of a
near-extremal BH is no longer a degree-two homogeneous function of charges \cite%
{Chattopadhyay:2021vog}, a consistent generalization of FD has been achieved
in \cite{Chattopadhyay:2022ycb} only at the price of transforming the
temperature $T$. The resulting map, which is no longer anti-involutive (but
nevertheless is analytical and unique), has been termed \textit{%
near-extremal (generalized}\footnote{%
Not to be confused with the \textit{general Freudenthal transformations}
(GFT's) treated in \cite{Borsten:2019xas}.}\textit{) FD} : two near-extremal
BHs, whose charges and (small) temperatures are connected by the
near-extremal FD, have the same entropy.\medskip

In this paper, we aim to further extend the notion of FD to the class of
extremal (stationary) \textit{rotating} BHs, within the same spirit as \cite%
{Chattopadhyay:2022ycb}. In \cref{sec:F_dual}, we show that in general the
definition of the Freudenthal dual for an extremal (asymptotically flat)
over-(or under-)rotating BH is \textit{not} consistent while keeping its
angular momentum fixed. To overcome this issue, in \cref{sec:GFD} we define
a - generally, non-anti-involutive - \textit{(generalized)} \textit{rotating
FD} (RFD) map: interestingly, within our formulation, RFD uniquely maps the
dyonic e.m. charges and the angular momentum $J$ of an extremal rotating BH
to the ones of another extremal rotating BH, while keeping the
Bekenstein-Hawking BH entropy fixed.\medskip

The plan of the paper is as follows. After a first, na\"{\i}ve generalizing
approach and some preliminary considerations in Sec. \ref{sec:F_dual}, we
present the generalization of FD to RFD in Sec. \ref{sec:GFD}, analyzing in
detail its non-rotating limit in Sec. \ref{sec:golden}. Then, after
presenting numerical evidence in Sec. \ref{sec:num}, we prove the uniqueness
of the RFD map in Sec. \ref{subsec:sextic}, and determine its analytic form
in Sec. \ref{sec:an}. Finally, in Sec. \ref{sec:underover} we investigate
the possibility of inducing transitions between the (stationary) under- and
over-rotating regimes by means of the RFD map, obtaining a \textit{no-go}
result. Final remarks and hints for further developments are given in the
concluding Sec. \ref{sec:conc}. Apps. \ref{app:solj} and \ref{app:proof},
containing details concerning the treatment given in Sec. \ref{sec:underover}%
, conclude the paper.


\section{\label{sec:F_dual}Freudenthal duality (FD) and rotating BHs}

In this section, we analyse the prospect of having an F-dual of an extremal,
over-(or under)-rotating BH, such that both of them have the same angular
momentum and entropy whereas their e.m. charges are related by FD.

Before proceeding further, we first briefly review the Freudenthal duality for extremal, non-rotating black holes. For such  black holes with  e.m. charges  collected into the symplectic vector $Q$ and entropy $S_0(Q)$, the Freudenthal duality acts non-trivially  on $Q$ as follows
\begin{eqnarray}\label{usualFD}
	Q\longmapsto \hat{Q}(Q) &:=& \Omega\frac{\partial S_0(Q)}{\partial Q}
\end{eqnarray} 

such that %
\begin{equation}
	S_{0}(Q)=S_{0}\left( \Omega \frac{\partial S_{0}\left( Q\right) }{\partial Q}%
	\right) .  \label{FD-inv}
\end{equation}
$\Omega$ is a symplectic matrix with $\Omega^T=-\Omega$ and $\Omega^2=- \mathbb{I}$. In other words, two non-rotating, extremal black holes with e.m. charge vectors $Q$ and $\hat{Q}$ related to each other by \eqref{usualFD} have the same value for their entropy.  The non-rotating, extremal entropy is a degree-two homogeneous function in the e.m. charge $Q$ i.e. 
\begin{equation}
	S_{0}\left( \lambda Q\right) =\lambda ^{2}S_{0}(Q),~\forall \lambda \in
	\mathbb{R},  \label{hom-2-Q}
\end{equation}%
and this property plays a fundamental role for the invariance of the entropy  under  \eqref{usualFD}.

It is worth noticing that, in this case, the transformation \eqref{usualFD} is anti-involutive, i.e. $\hat{\hat{Q}}=-Q.$

The study of the Freudenthal duality was further extended in a non-anti-involutive way for the near-extremal, non-rotating black holes in \cite{Chattopadhyay:2021vog, Chattopadhyay:2022ycb}.  Due to the presence of the temperature $T$ in the form,  the entropy of a near-extremal, non-rotating black hole is no more a degree-two homogeneous function of e.m. charges. Hence, the obvious way of charge transformation  following \eqref{usualFD}, i.e. $\hat{Q}(Q;T) := \Omega \frac{\partial S_{NE}(Q;T)}{
\partial Q}$, called as "on-shell" FD does not keep the near-extremal entropy invariant. In spite of this,  it has been shown that two different near-extremal, non-rotating black holes with charge and temperature respectively $(Q, T)$ and $(\hat{Q}, T+\delta T)$ have the same entropy when

\begin{eqnarray}\label{nearextFD}
	Q\longmapsto \hat{Q}(Q; T+\delta T) &:=& \Omega\frac{\partial S_{NE}(Q; T+\delta T)}{\partial Q}
\end{eqnarray} 
with a unique solution for $\delta T$.

With this brief recap, we now establish the Freudenthal duality for  extremal, rotating black holes.

After \cite{Astefanesei:2006dd,Ferrara:2008hwa}, the Bekenstein-Hawking
entropy for an extremal, (asymptotically flat) under-rotating Kerr-Newman BH  with angular momentum $J$ and e.m. charge $Q$
is given by
\begin{eqnarray}
S_{\text{under}}\left( Q,J\right) &=&\sqrt{S_{0}^{2}\left( Q\right) -J^{2}};
\label{enunder} \\
S_{0}^{2}\left( Q\right) -J^{2} &>&0,  \label{2}
\end{eqnarray}%
whereas the entropy for an extremal, (asymptotically flat) over-rotating
Kerr-Newman BH is\footnote{%
Clearly, only the under-rotating class (\ref{enunder})-(\ref{2}) of
solutions have a well-defined non-rotating limit.}
\begin{eqnarray}
S_{\text{over}}\left( Q,J\right) &=&\sqrt{J^{2}-S_{0}^{2}\left( Q\right) };
\label{enover} \\
J^{2}-S_{0}^{2}\left( Q\right) &>&0.  \label{bis}
\end{eqnarray}%
	In general, within the semi-classical supergravity approximation, the e.m.
	charges  take real
	values, and so do the angular momentum $J$ and the non-rotating entropy $%
	S_{0}(Q)$, which are further constrained to be non-negative and strictly
	positive, respectively: $J\in \mathbb{R}^{+}$, $S_{0}\in \mathbb{R}_{0}^{+}$.%

Following the usual formulation of FD \cite{Ferrara:2011gv}, the action of
the \textquotedblleft on-shell\textquotedblright\ FD map on $Q$ reads
\begin{equation}
Q\longmapsto \hat{Q}\left( Q,J\right) :=\Omega \frac{\partial S_{\text{%
under(over)}}\left( Q,J\right) }{\partial Q}=\pm \frac{S_{0}\left( Q\right)
}{\sqrt{\left\vert S_{0}^{2}\left( Q\right) -J^{2}\right\vert }}\Omega \frac{%
\partial S_{0}\left( Q\right) }{\partial Q},  \label{canf}
\end{equation}%
with $\Omega $ denoting the symplectic invariant structure of the e.m.
charge representation space ($\Omega ^{T}=-\Omega $ and $\Omega ^{2}=-%
\mathbb{I}$); the \textquotedblleft $+$\textquotedblright\ and
\textquotedblleft $-$\textquotedblright\ signs correspond to the
under-rotating and over-rotating cases respectively. Thus, a na\"{\i}ve
generalization of FD (leaving $J$ fixed), in these cases, can be formulated
as a map acting on $Q$ as (\ref{canf}), such that
%
\begin{equation}
S_{\text{under(over)}}\left( Q,J\right) =S_{\text{under(over)}}\left( \Omega
\frac{\partial S_{\text{under(over)}}\left( Q,J\right) }{\partial Q}%
,J\right) .  \label{eq:pre-condd}
\end{equation}%
Using \eqref{enunder} and \eqref{enover}, one can show that for both the
under-rotating and over-rotating extremal Kerr-Newman BHs, the conditions
for the existence of a Freudenthal dual boils down to solving a very simple
algebraic equation, namely \footnote{Note that $S_0\equiv S_0 (Q) $.}
\begin{equation}
S_{0}^{2}J^{2}\left( J^{2}-2S_{0}^{2}\right) =0.  \label{eq:condd}
\end{equation}

\subsection{Under-rotating}

For an under-rotating BH, $J^{2}-S_{0}^{2}(Q)<0$, and therefore $%
J^{2}-2S_{0}^{2}\left( Q\right) <0$ always. In this case, the only possible
solution of \eqref{eq:condd} is $J=0$. This is a trivial solution, since in
such a limit \eqref{eq:pre-condd} reduces to the usual (and anti-involutive)
notion of FD for a non-rotating, extremal BH \cite{Ferrara:2011gv}.

\subsection{\label{sec:over}Over-rotating}

For an over-rotating BH, $J^{2}-S_{0}^{2}\left( Q\right) >0$, and this rules
out the possibility of the solution $J=0$ to \eqref{eq:condd}. Still, one
could define the F-dual of an over-rotating extremal Kerr-Newman BH, when
the condition \eqref{eq:condd} is met as
\begin{equation}
J^{2}=2S_{0}^{2}\left( Q\right) \Rightarrow S_{\text{over}}\left( Q,J\right)
=S_{0}(Q).
\end{equation}%
Thus, one can conceive the starting extremal BH as a non-rotating BH, for
which having a F-dual is obvious \cite{Ferrara:2011gv}.

\section{\label{sec:GFD}Generalized rotating Freudenthal duality (RFD)}

We have just observed that, by the usual notion of FD, one cannot map two
extremal rotating BHs to each other. In this section, we go beyond the usual
definition of FD and try to map two extremal rotating BHs (both under- or
over-rotating) with \textit{different} angular momenta.

Namely, motivated by the approach carried out in \cite{Chattopadhyay:2022ycb}
for a near-extremal BH with the temperature $T$, here we define the
transformation of the dyonic e.m. BH charges as follows :
%
\begin{eqnarray}
Q &\rightarrow &\hat{Q}(Q,J+\delta J):=\Omega \frac{\partial S_{\text{%
under(over)}}\left( Q,J+\delta J\right) }{\partial Q}  \notag \\
&=&\pm \frac{S_{0}\left( Q\right) }{\sqrt{\left\vert S_{0}^{2}\left(
Q\right) -\left( J+\delta J\right) ^{2}\right\vert }}\Omega \frac{\partial
S_{0}\left( Q\right) }{\partial Q},  \label{Q-transf}
\end{eqnarray}%
Where the branch \textquotedblleft $+$\textquotedblright\ and
\textquotedblleft $-$\textquotedblright\ represents the under-rotating ($%
S_{0}^{2}\left( Q\right) >\left( J+\delta J\right) ^{2}$) and over-rotating (%
$S_{0}^{2}\left( Q\right) <\left( J+\delta J\right) ^{2}$) cases
respectively, such that the Bekenstein-Hawking BH entropy remains invariant,
i.e.
\begin{equation}
S_{\text{under(over)}}\left( Q,J\right) =S_{\text{under(over)}}\left( \hat{Q}%
(Q,J+\delta J),J+\delta J\right) .  \label{enequal}
\end{equation}%
Therefore, the set of transformations%
\begin{equation}
\text{RFD}:\left\{
\begin{array}{l}
Q\rightarrow \hat{Q}(Q,J+\delta J); \\
~ \\
J\rightarrow J+\delta J;%
\end{array}%
\right.  \label{RFD}
\end{equation}%
define the \textit{(generalized) rotating FD} (RFD) map. By using %
\eqref{enunder}, \eqref{enover}, for both under- and over- rotating BHs, the
condition \eqref{enequal} boils down to the following sextic algebraic
inhomogeneous equation in the variation $\delta J$ of the angular momentum :
\begin{equation}
a_{1}\left( \delta J\right) ^{6}+a_{2}\left( \delta J\right)
^{5}+a_{3}\left( \delta J\right) ^{4}+a_{4}\left( \delta J\right)
^{3}+a_{5}\left( \delta J\right) ^{2}+a_{6}\delta J+a_{7}=0,
\label{eq:sextic1}
\end{equation}%
where each coefficients is a function of $J$ and $S_{0}=S_{0}(Q)$,
\begin{eqnarray}
a_{1} &=&1;  \notag \\
a_{2} &=&6J;  \notag \\
a_{3} &=&14J^{2}-S_{0}^{2};  \notag \\
a_{4} &=&4\left( 4J^{2}-S_{0}^{2}\right) J;  \notag \\
a_{5} &=&-S_{0}^{4}+9J^{4}-4S_{0}^{2}J^{2};  \notag \\
a_{6} &=&-2\left( S_{0}^{4}-J^{4}\right) J;  \notag \\
a_{7} &=&\left( J^{2}-2S_{0}^{2}\right) S_{0}^{2}J^{2}.  \label{cond-6}
\end{eqnarray}%
In order to find out two rotating extremal BHs with different angular
momenta but the same entropy, having their dyonic charges related by RFD (%
\ref{RFD}), we need to look for a real solution of \eqref{eq:sextic1},
namely for $\delta J=\delta J(J,S_{0})\in \mathbb{R}$ such that the
transformed angular momentum reads $\hat{J}:=J+\delta J\in \mathbb{R}^{+}$.

\subsection{\label{sec:golden}Non-rotating limit of RFD : the spurious,
\textquotedblleft golden\textquotedblright\ branch}

Before dealing with a detailed analysis of the roots of the algebraic
equation \eqref{eq:sextic1}, let us investigate the limit $J\rightarrow
0^{(+)}$ of the set of solutions to (\ref{eq:sextic1}). In this limit, which
is well-defined only for the under-rotating class of solutions, $\hat{J}%
\rightarrow \delta J^{(+)}\in \mathbb{R}^{+}$, and Eq. (\ref{eq:sextic1})
reduces to
\begin{equation}
(\delta J)^{6}-S_{0}^{2}(\delta J)^{4}-S_{0}^{4}(\delta J)^{2}=0,
\label{eq:sexticJ0}
\end{equation}%
which consistently admits the solution $\delta J=0$. This is no surprise,
since in the $J\rightarrow 0^{(+)}$ limit, RFD simplifies down to its usual,
non-rotating definition (namely, to FD, \cite{Ferrara:2011gv}).

Interestingly, other two real solutions to the cubic algebraic homogeneous
equation (in $(\delta J)^{2}$) (\ref{eq:sexticJ0}) exist. In fact, Eq. %
\eqref{eq:sexticJ0} can be solved by two more real roots\footnote{%
There exist also two purely imaginary roots with $\delta
J^{2}=S_{0}^{2}\left( {\frac{1-\sqrt{5}}{2}}\right) $, which we ignore.}

\begin{equation}
\delta J_{\pm }:=\pm \sqrt{\phi }\,S_{0}\left( Q\right) ,  \label{golden}
\end{equation}%
with $\phi :={\frac{1+\sqrt{5}}{2}}$ being the so-called \textit{golden ratio%
} (see e.g. \cite{TheFibonacciSequenceandtheGoldenRatio}).
Since $\hat{J}\rightarrow \delta J^{(+)}\in \mathbb{R}^{+}$, only $\delta
J_{+}$ in (\ref{golden}) has a sensible physical meaning. Correspondingly,
within the $J\rightarrow 0^{+}$ (i.e., non-rotating) limit, the unique
non-vanishing, physically sensible solution for the angular momentum
transformation reads
\begin{equation}
\hat{J}_{\text{golden}}:=\delta J_{+}=\sqrt{\phi }S_{0}\left( Q\right) =%
\sqrt{\frac{1+\sqrt{5}}{2}}S_{0}(Q).  \label{eq:delJat0}
\end{equation}%
This analysis shows the existence, in the limit $J\rightarrow 0^{+}$, of a
spurious, \textquotedblleft golden\textquotedblright\ branch of RFD, which
in the non-rotating limit does not reduce to the usual FD, but rather it
allows to map a non-rotating extremal BH to an under-rotating (stationary)
one; this can be depicted as
\begin{equation}
\underset{\text{static~extremal BH}}{\left\{
\begin{array}{l}
S=S_{0}\left( Q\right) \\
J=0%
\end{array}%
\right. }~\overset{\text{RFD}_{J\rightarrow 0^{+}\text{,~golden}}}{%
\longrightarrow }\underset{\text{under-rotating~extremal~BH}}{\left\{
\begin{array}{l}
S_{\text{under}}=S_{\text{under}}\left( \hat{Q}_{\text{golden}},\hat{J}_{%
\text{golden}}\right) \\
J=\hat{J}_{\text{golden}}%
\end{array}%
\right. }~,  \label{depict}
\end{equation}%
where $\hat{J}_{\text{golden}}$ is defined by (\ref{eq:delJat0}), $\hat{Q}_{%
\text{golden}}$ is defined by%
\begin{eqnarray}
\hat{Q}_{\text{golden}} &:=&\Omega \frac{\partial S_{\text{over}}\left( Q,%
\hat{J}_{\text{golden}}\right) }{\partial Q}=-\frac{S_{0}\left( Q\right) }{%
\sqrt{\hat{J}_{\text{golden}}^{2}-S_{0}^{2}\left( Q\right) }}\Omega \frac{%
\partial S_{0}\left( Q\right) }{\partial Q}  \notag \\
&=&-\frac{S_{0}\left( Q\right) }{\sqrt{\phi S_{0}^{2}(Q)-S_{0}^{2}\left(
Q\right) }}\Omega \frac{\partial S_{0}\left( Q\right) }{\partial Q}=-\frac{1%
}{\sqrt{\phi -1}}\Omega \frac{\partial S_{0}\left( Q\right) }{\partial Q}
\notag \\
&=&-\sqrt{\phi }\Omega \frac{\partial S_{0}\left( Q\right) }{\partial Q},
\label{Qhat}
\end{eqnarray}%
and RFD$_{J\rightarrow 0^{+}\text{,~golden}}$ denotes such a spurious,
\textquotedblleft golden\textquotedblright\ branch of the non-rotating limit
of RFD (\ref{RFD}) :%
\begin{equation}
\text{RFD}_{J\rightarrow 0^{+}\text{,~golden}}:\left\{
\begin{array}{l}
Q\rightarrow \hat{Q}_{\text{golden}}=-\sqrt{\phi }\Omega \frac{\partial
S_{0}\left( Q\right) }{\partial Q}; \\
~ \\
J=0\rightarrow \hat{J}_{\text{golden}};%
\end{array}%
\right.  \label{RFD-J=0}
\end{equation}%
Note that in the last step of (\ref{Qhat}) we used the crucial property of
the golden ratio $\phi $, namely%
\begin{equation}
\phi -1=\frac{1}{\phi }.  \label{cruc}
\end{equation}%
Consistently, the total entropy is preserved by the map $\left( \text{RFD}%
\right) _{J\rightarrow 0^{+}\text{,~golden}}$ (\ref{RFD-J=0}), because%
\begin{eqnarray}
S_{\text{under}}\left( \hat{Q}_{\text{golden}},\hat{J}_{\text{golden}%
}\right) &=&\sqrt{S_{0}^{2}\left( \hat{Q}_{\text{golden}}\right) -\hat{J}_{%
\text{golden}}^{2}}  \notag \\
&=&\sqrt{S_{0}^{2}\left( -\sqrt{\phi }\Omega \frac{\partial S_{0}\left(
Q\right) }{\partial Q}\right) -\phi S_{0}^{2}\left( Q\right) }  \notag \\
&=&\sqrt{\phi ^{2}S_{0}^{2}\left( \Omega \frac{\partial S_{0}\left( Q\right)
}{\partial Q}\right) -\phi S_{0}^{2}\left( Q\right) }  \notag \\
&=&\sqrt{\phi ^{2}-\phi }S_{0}\left( Q\right) =S_{0}\left( Q\right),
\label{!}
\end{eqnarray}%
where in the last step the crucial property (\ref{cruc}) has been used
again. Moreover, in achieving (\ref{!}), we have also exploited two crucial
properties of the static extremal BH entropy $S_{0}(Q)$, namely its
homogeneity of degree two in the e.m. charges, and its invariance under the Freudenthal duality for static extremal BHs, as given by \eqref{hom-2-Q} and \eqref{FD-inv} respectively.%

Thus, the two extremal BHs in the l.h.s and r.h.s of (\ref{depict}) have the
same Bekenstein-Hawking entropy, preserved by the map $\left( \text{RFD}%
\right) _{J\rightarrow 0^{+}\text{,~golden}}$ (\ref{RFD-J=0}).

\paragraph*{Remark 1}

It is worth remarking that the stationary extremal BH in the r.h.s. of (\ref%
{depict}), namely the image of an extremal, static BH under the map RFD$%
_{J\rightarrow 0^{+}\text{,~golden}}$ (\ref{RFD-J=0}), is necessarily
\textit{under-rotating}, because the non-rotating limit $J\rightarrow 0^{+}$
is well-defined \textit{only} in the under-rotating case. On the other hand,
the BH entropy entering the definition of $\hat{Q}_{\text{golden}}$ (\ref%
{Qhat}) is necessarily of the \textit{over-rotating} type, since%
\begin{equation}
\phi >1\Rightarrow \hat{J}_{\text{golden}}=\sqrt{\phi }S_{0}\left( Q\right)
>S_{0}\left( Q\right) \Leftrightarrow \hat{J}_{\text{golden}%
}^{2}-S_{0}^{2}\left( Q\right) >0,
\end{equation}%
which pertains to the over-rotating case.

\paragraph*{Remark 2}

In a sense, the use of the spurious, \textquotedblleft
golden\textquotedblright\ branch $\left( \text{RFD}\right) _{J\rightarrow
0^{+}\text{,~golden}}$ (\ref{RFD-J=0}) of the map RFD (\ref{RFD}) can be
regarded as a kind of solution-generating technique, which generates an
under-rotating, stationary extremal BH from a non-rotating, static BH, while
keeping the Bekenstein-Hawking entropy fixed. Indeed, while both such BHs
are asymptotically flat, their near-horizon geometry changes under $\left(
\text{RFD}\right) _{J\rightarrow 0^{+}\text{,~golden}}$ :%
\begin{equation}
\underset{\text{static~extremal BH~\cite{Kunduri:2007vf}}}{AdS_{2}\otimes
S^{2}}~\overset{\left( \text{RFD}\right) _{J\rightarrow 0^{+}\text{,~golden}}%
}{\longrightarrow }\underset{\text{under-rotating~extremal~BH \cite%
{Bardeen:1999px}}}{AdS_{2}\otimes S^{1}}~.  \label{depict-nhg}
\end{equation}

\paragraph*{Remark 3}

The expression of the Bekenstein-Hawking entropy of under-rotating resp.
over-rotating stationary extremal BHs, respectively given by (\ref{enunder}%
)-(\ref{2}) and (\ref{enover})-(\ref{bis}), is suggestive of a
representation of the two fundamental quantities $S_{0}$ and $J$
characterizing such BHs in terms of elements of the \textit{split complex}
(also named \textit{hypercomplex}) numbers $\mathbb{C}_{s}$ (see e.g. \cite%
{GaussianParabolicandHyperbolicNumbers}), i.e. respectively as%
\begin{eqnarray}
Z &:=&S_{0}+\mathbf{i}J\in \mathbb{C}_{s}~\Rightarrow S_{\text{under}%
}=\left\vert Z\right\vert :=\sqrt{Z\bar{Z}}=\sqrt{S_{0}^{2}-J^{2}}; \\
Y &:=&\mathbf{i}Z=J+\mathbf{i}S_{0}\in \mathbb{C}_{s}~\Rightarrow S_{\text{%
over}}=\left\vert Y\right\vert :=\sqrt{-Z\bar{Z}}=\sqrt{J^{2}-S_{0}^{2}},
\end{eqnarray}%
where $\mathbf{i}$ denotes the split imaginary unit ($\mathbf{i}^{2}=1$, not
to be confused with $\pm 1$), and the bar stands for the hypercomplex
conjugation, defined as $\bar{Z}:=S_{0}-\mathbf{i}J$.

Thus, any map acting on $S_{0}$ and $J$, as the RFD map (\ref{RFD}) itself,
can be represented as acting on the hypercomplex (analogue of the)
Argand-Gauss plane, denoted by $\mathbb{C}_{s}\left[ S_{0},J\right] $ resp. $%
\mathbb{C}_{s}\left[ J,S_{0}\right] $ for under- resp. over-rotating BHs.
Since $\mathbb{C}_{s}$ is not a field (because it contains zero divisors due
to its split nature), there are no split analogues of the \textquotedblleft
wild\textquotedblright\ automorphisms \cite{Kestelman} in $\mathbb{C}_{s}$,
but rather only four discrete fundamental automorphisms, namely $\mathbb{I}$%
, $-\mathbb{I}$, $\mathbf{C}$ and $-\mathbf{C}$, where $\mathbb{I}$ and $%
\mathbf{C}$ respectively denote the identity map and the aforementioned
conjugation map on $\mathbb{C}_{s}\left[ S_{0},J\right] $ or $\mathbb{C}_{s}%
\left[ J,S_{0}\right] $; interestingly, since $S_{0}\in \mathbb{R}_{0}^{+}$
and $J\in \mathbb{R}^{+}$, none (but, trivially, $\mathbb{I}$) of such
discrete automorphisms are physically allowed, and the physically sensible
quadrant of $\mathbb{C}_{s}\left[ S_{0},J\right] $ resp. $\mathbb{C}_{s}%
\left[ J,S_{0}\right] $ is only the first one (with the $J$ axis excluded).

In particular, the RFD map acts on any $Z\in \mathbb{C}_{s}\left[ S_{0},J%
\right] $ resp. $Y\in \mathbb{C}_{s}\left[ J,S_{0}\right] $ as a
norm-preserving transformation, because, by definition, it preserves the
Bekenstein-Hawking entropy $S$, as given by the defining condition (\ref%
{enequal}). Thus, considering a under- resp. over-rotating (stationary,
asymptotically flat) extremal BH with Bekenstein-Hawking entropy $S=\mathcal{%
S}\in \mathbb{R}^{+}$, its RFD-dual extremal BH will belong to the arc of
the hyperbola $\left\vert Z\right\vert ^{2}=S_{0}^{2}-J^{2}=\mathcal{S}$
resp. $\left\vert Y\right\vert ^{2}=J^{2}-S_{0}^{2}=\mathcal{S}$ within the
first quadrant (with the $J$ axis excluded) of the hypercomplex Argand-Gauss
plane.

In the non-rotating limit, in light of the treatment of Sec. \ref{sec:golden}%
, the action of the RFD map (\ref{RFD}) on a non-rotating extremal BH
(represented as a point of the $S_{0}$ axis - with its origin excluded - in
the hyper-Argand-Gauss plane $\mathbb{C}_{s}\left[ S_{0},J\right] $, thus
with coordinates $\left( S_{0},0\right) $) may be nothing but the identity $%
\mathbb{I}$ (in the usual, non-rotating branch of RFD$_{J\rightarrow 0^{+}}$%
) or a point with coordinates $\left( \phi S_{0},\sqrt{\phi }S_{0}\right) $
along the arc of hyperbola defined as $\mathcal{H}:=\left\{ Z\in \mathbb{C}%
_{s}\left[ S_{0},J\right] :\left\vert Z\right\vert ^{2}=S_{0}^{2}\right\} $
(in the spurious, \textquotedblleft golden\textquotedblright\ branch RFD$%
_{J\rightarrow 0^{+}\text{,~golden}}$ of RFD$_{\text{J}\rightarrow 0^{+}}$,
discussed above).


\section{\label{sec:num}Numerical interlude}

Before carrying out an analytical study of the roots of the algebraic Eq. (%
\ref{eq:sextic1}), we want to present numerical evidence that for $J>0$
(thus going beyond the non-rotating limit), only two roots, out of the six
roots of the sextic inhomogeneous algebraic Eq. (\ref{eq:sextic1}) (which
exist in the algebraically closed field of complex numbers $\mathbb{C}$, by
the fundamental theorem of algebra), are real roots. Moreover, as shown in
the plots of Fig. 1, out of such two real roots, named $\delta J_{1}$ and $%
\delta J_{2}$, only one, say $\delta J_{2}$, is consistent with the
requirement of physical soundness, namely with the condition that $\hat{J}%
:=J+\delta J\geqslant 0$. Fig. 1(a) shows the plot\footnote{%
In the same plot, the consideration of the under-rotating or of the
over-rotating case just affects the range of the plot, while not affecting
the uniqueness of the real (and physically sensible) root.}of $J+\delta J$
vs. $J$ for fixed entropy, whereas Fig. 1(b) shows the plot of $\delta J$
vs. $J$, for fixed entropy for the two aforementioned real roots of Eq. (\ref%
{eq:sextic1}). It is worth here remarking that $\delta J_{2}$ actually goes
into \eqref{eq:delJat0} in the limit $J\rightarrow 0^{+}$, as expected.
\begin{figure}[tbph]
\begin{subfigure}{0.48\textwidth}
		\centering
		\includegraphics[width=\textwidth]{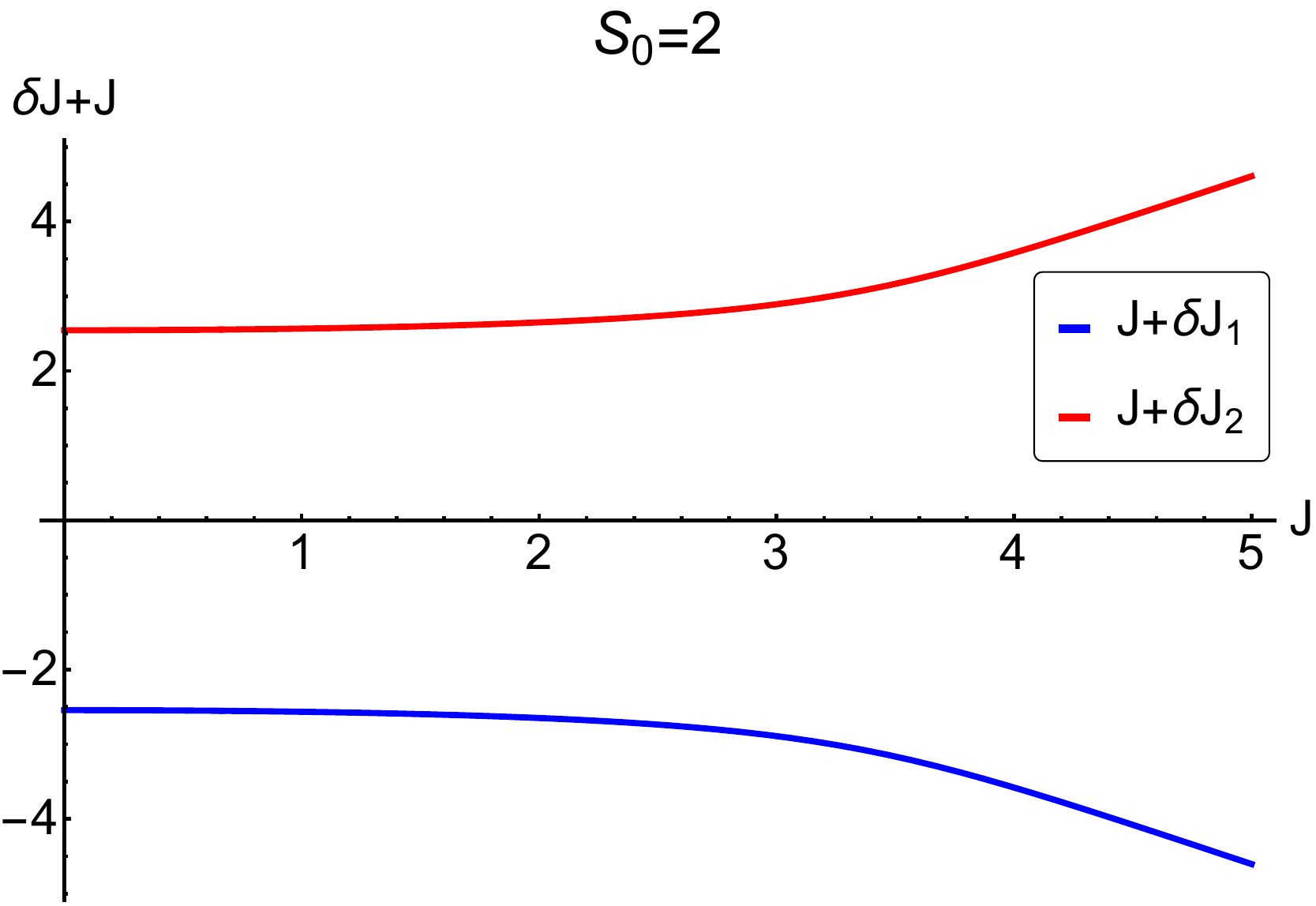}
		\caption{}
	\end{subfigure}\hfill
\begin{subfigure}{0.48\textwidth}
		\centering
		\includegraphics[width=\textwidth]{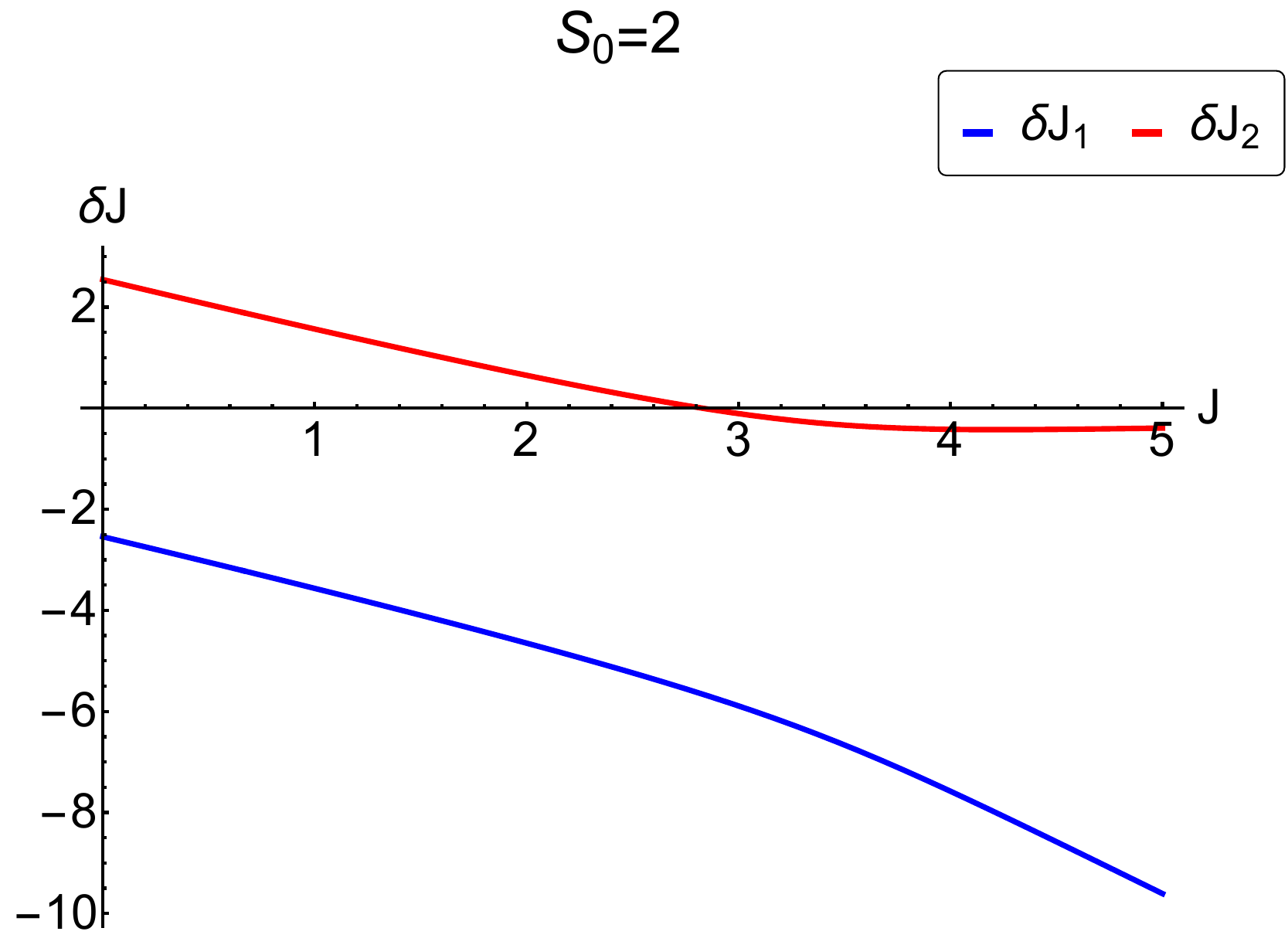}
		\caption{}
	\end{subfigure}
\caption{The two real roots of (\protect\ref{eq:sextic1})}
\label{fig:numerics}
\end{figure}

\section{\label{subsec:sextic}Uniqueness of RFD}

Recalling the definition $\hat{J}:=J+\delta J\in \mathbb{R}^{+}$ of the
(physically sensible) RFD-dual angular momentum as, the sextic algebraic Eq.
(\ref{eq:sextic1})-(\ref{cond-6}) can be rewritten as
\begin{equation}
\hat{J}^{6}-\hat{J}^{4}(J^{2}+S_{0}^{2})-\hat{J}%
^{2}S_{0}^{2}(S_{0}^{2}-2J^{2})-J^{2}S_{0}^{4}=0.  \label{jsextic}
\end{equation}%
By setting $x:=\hat{J}^{2}$, \eqref{jsextic} becomes an inhomogeneous cubic
algebraic equation,
\begin{equation}
x^{3}-x^{2}(J^{2}+S_{0}^{2})-x(S_{0}^{2}-2J^{2})S_{0}^{2}-J^{2}S_{0}^{4}=0.
\label{eq:cubic1}
\end{equation}%
In order to analyse the roots of (\ref{eq:cubic1}), we consider the
following algebraic curve\footnote{%
The $y$ intercept, $f(0)=-J^{2}S_{0}^{4}$, is always negative, or zero only
at $J=0$.}
\begin{equation}
f(x):=x^{3}-x^{2}(J^{2}+S_{0}^{2})-x(S_{0}^{2}-2J^{2})S_{0}^{2}-J^{2}S_{0}^{4},
\label{eq:fcurve}
\end{equation}%
and look for the \textit{turning points} on $x$ vs. $y=f(x)$, indeed,
counting the number of times $f(x)$ crosses the $x$-axis provides the number
of real roots.

In general, a cubic equation has either one or three real roots, and \textit{%
at most} two turning points. The collective information about the locations
of the turning points and the $x$-intercept of $f(x)$ or the position of $%
f(0)$ provides us with a clear picture of the situation. In order to compute
the number of turning points or the points where the slope goes to zero, one
has to solve
\begin{equation*}
{\frac{df\left(x\right) }{dx}}=(S_{0}^{2}-x)(2J^{2}-S_{0}^{2}-3x)=0,
\end{equation*}%
which generates the following turning points $T_{i}:=(x_{i},f(x_{i}))$, $%
i=1,2$ :
\begin{eqnarray}
T_{1} &:=&\left( S_{0}^{2},-S_{0}^{6}\right) ,\quad T_{2}:=\left( {\frac{1}{3%
}}(2J^{2}-S_{0}^{2}),{\frac{\mathcal{A}}{27}}\right) ,  \label{turningpt} \\
\mathcal{A} &:=&-4(J^{2}-2S_{0}^{2})^{3}-27S_{0}^{6}.  \label{eq:Aeq}
\end{eqnarray}%
As $S_{0}>0$, $T_{1}$ always lies in the fourth quadrant of the $x-y$ plane.
From the values of $f(x)$ defined in (\ref{eq:fcurve}) at both the turning
points $T_{1}$ and $T_{2}$, one can conclude the number of real roots for
the cubic equation $f(x)=0$ (\ref{eq:cubic1}). For further insight, one can
look at the second derivatives at the turning points,
\begin{equation}
\mathcal{S}_{1}:=\left. {\frac{d^{2}f\left( x\right) }{dx^{2}}}\right\vert
_{x=S_{0}^{2}}=-2(J^{2}-2S_{0}^{2}),\quad \mathcal{S}_{2}:=\left. {\frac{%
d^{2}f}{dx^{2}}}\right\vert _{x=1/3(2J^{2}-S_{0}^{2})}=2(J^{2}-2S_{0}^{2}),
\end{equation}%
implying that the convexity of the function $y=f(x)$ always remains opposite
at $T_{1}$ and $T_{2}$. The case $J^{2}=2S_{0}^{2}$, which can only arise
for over-rotating BHs, will be discussed further below.

The same information can be derived from the discriminant of the cubic
equation \eqref{eq:fcurve}, which can be computed to read
\begin{equation}
\Delta =S_{0}^{6}\,\mathcal{A}.  \label{discriminant}
\end{equation}%
For any cubic equation, if the discriminant $\Delta >0$, there are three
real roots, whereas if $\Delta <0$, there is only one real root. To
determine the number of real roots of $f(x)=0$, one then needs to check the
sign of either the discriminant or the value of $f(x)$ at the turning
points. Below, we will consider all possible cases.

\subsection{$J=0$ (non-rotating)}

We start and consider the simplest, i.e. the non-rotating, case: $J=0$. From %
\eqref{turningpt}, we find the following turning points :
\begin{equation}
T_{1}=\left\{ S_{0}^{2},-S_{0}^{6}\right\} ,\quad T_{2}=\left\{ -{\frac{%
S_{0}^{2}}{3}},{\frac{5\,S_{0}^{6}}{27}}\right\} .
\end{equation}%
Thus, $T_{2}$ always belongs to the second quadrant, with $\mathcal{S}_{1}>0$
and $\mathcal{S}_{2}<0$. Consequently, there exist $3$ real roots, as shown
in Fig 2(a). This also can be understood by computing that $\Delta
=5S_{0}^{12}>0$. Out of the resulting three real roots, one is zero, as $%
f(0)=0$, hence corresponding to the usual (and anti-involutive) notion of FD
for a non-rotating, extremal BH \cite{Ferrara:2011gv}. We also observe that $%
T_{2}$ has a negative $x$ coordinate, whereas $T_{1}$ has a positive one :
this implies that among the two non-zero real roots, one is positive and the
other is negative. Since $x:=\hat{J}^{2}=(J+\delta J)^{2}$, only the
positive real root is a physically sensible choice: as discussed in Sec. \ref%
{sec:golden}, it corresponds to RFD mapping a non-rotating, static extremal
BH to an under-rotating, stationary extremal BH. In order to further clarify
the locations of the turning point, as an example in Fig. 2(a) we plot $%
y=f(x)$ when $J=0$ and $S_{0}=5$.
\begin{figure}[tbph]
\begin{subfigure}{0.48\textwidth}
		\centering
\includegraphics[width=\textwidth]{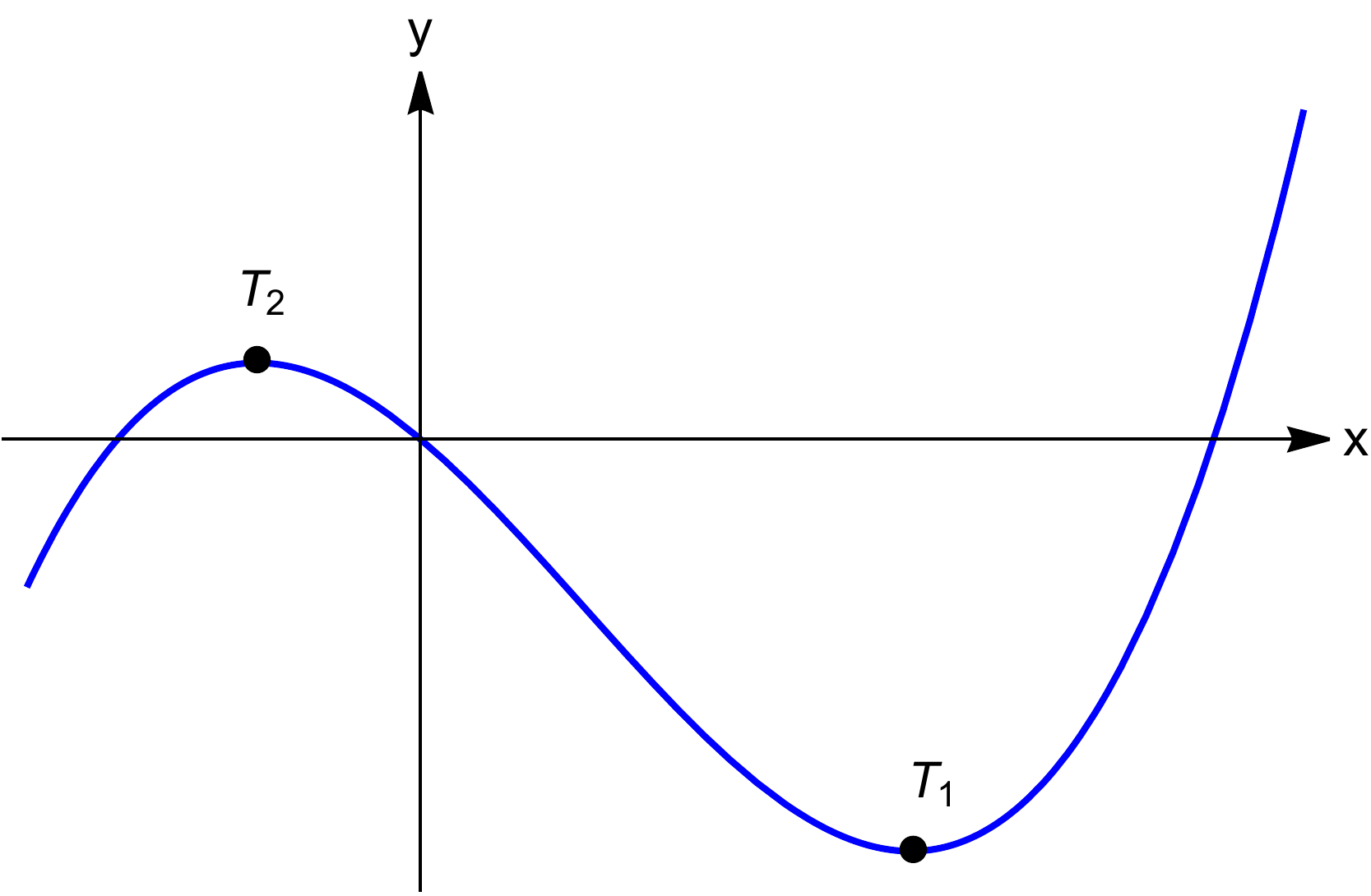}
		\caption{$J=0$.}
\label{fig:cas1J0}
	\end{subfigure}\hfill
\begin{subfigure}{0.48\textwidth}
		\centering
    \includegraphics[width=\linewidth]{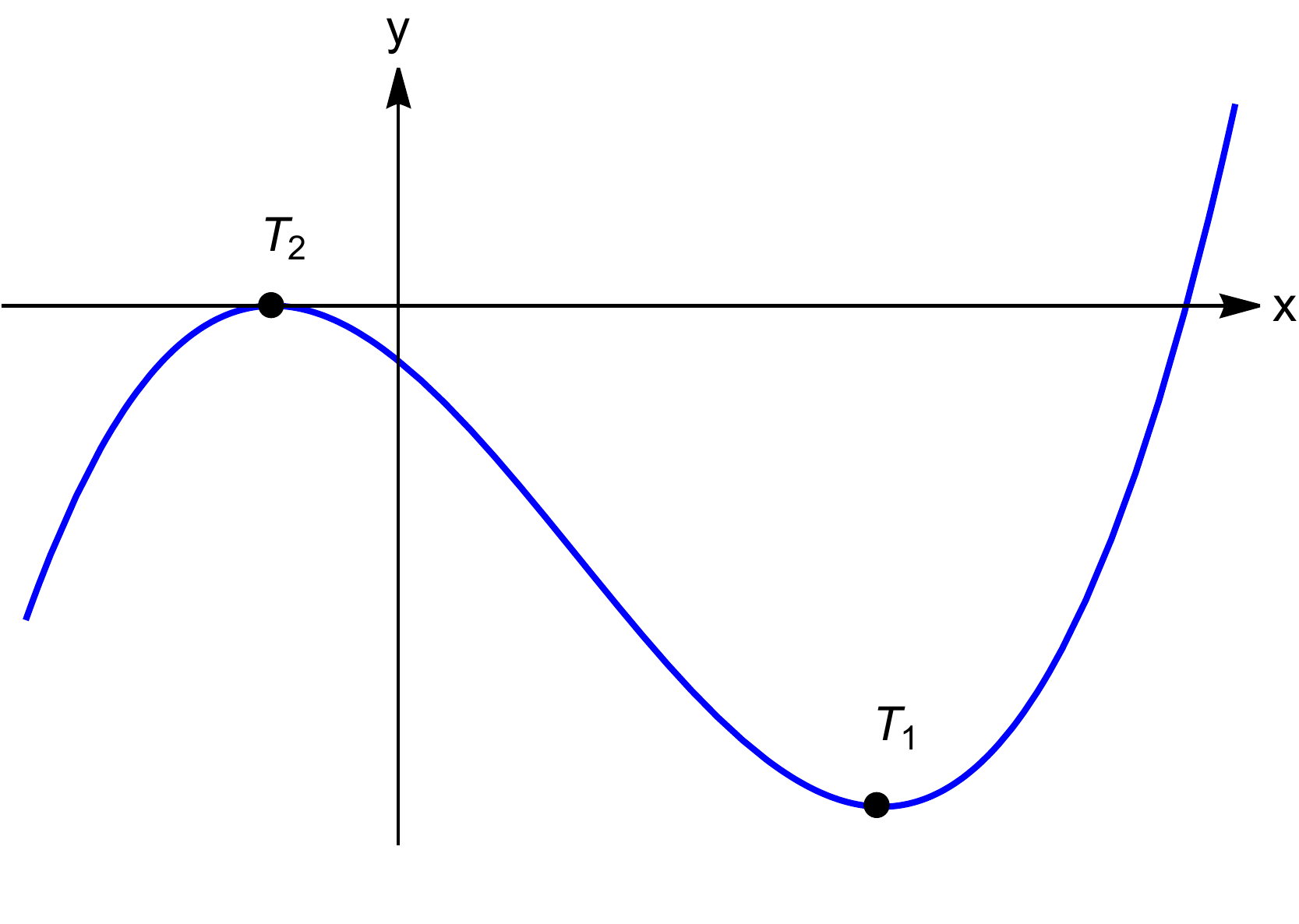}
		\caption{$J\neq 0$, $\cA=0$.}
\label{fig:cas2}
	\end{subfigure}
\caption{Typical plots for sections 5.1 and 5.2.}
\end{figure}

\subsection{$J\neq 0$, $\mathcal{A}=0$ (under-rotating)}

Next, we consider $J>0$, but $\mathcal{A}=0$. In this case, $\Delta =0$, and
therefore there are multiple real roots; since all the coefficients of the
cubic equation (\ref{eq:cubic1}) are real, then all roots are real. From %
\eqref{eq:Aeq}, we find that $\mathcal{A}=0\Rightarrow J^{2}=S_{0}^{2}%
\mathbf{g}$, where $\mathbf{g}:=\left( 2-{\frac{3}{2^{\frac{2}{3}}}}\right)
\approx 0.110118.$ Thus, $S_{0}^{2}-J^{2}=\left( 1-\mathbf{g}\right)
S_{0}^{2}>0$, implying that this is an under-rotating case. The turning
points read
\begin{equation}
T_{1}=\left\{ S_{0}^{2},-S_{0}^{6}\right\} ,\quad T_{2}=\left\{ {\frac{1}{3}}%
S_{0}^{2}(2\mathbf{g}-1),0\right\} .
\end{equation}%
Since there are two different turning points, again with $\mathcal{S}_{1}>0$
and $\mathcal{S}_{2}<0$, there exist two real roots, as depicted in Fig.
2(b) in the case in which $S_{0}=5$. Inserting the actual value of $\mathbf{g%
}$, we observe that $T_{2}$ has a vanishing $y=f(x)$ coordinate, but it has
a negative $x$ coordinate, and thus it must be discarded. On the other hand,
$T_{1}$ lies on the positive $x$ axis, implying the existence of only one
strictly positive real root, which is the physically acceptable one. By
setting $J^2=S^2_0 \mathbf{g}$ and using $\mathbf{g}\approx 0.110118$, one
can solve for the $f(x)=0$ and find that the multiplicity arises at the
negative real root, leaving a single positive real root.

\subsection{$J\neq 0$, $\mathcal{A}>0$ (under-rotating)\label{sec:this}}

\begin{wrapfigure}[11]{r}{0.45\textwidth}
\includegraphics[width=\linewidth]{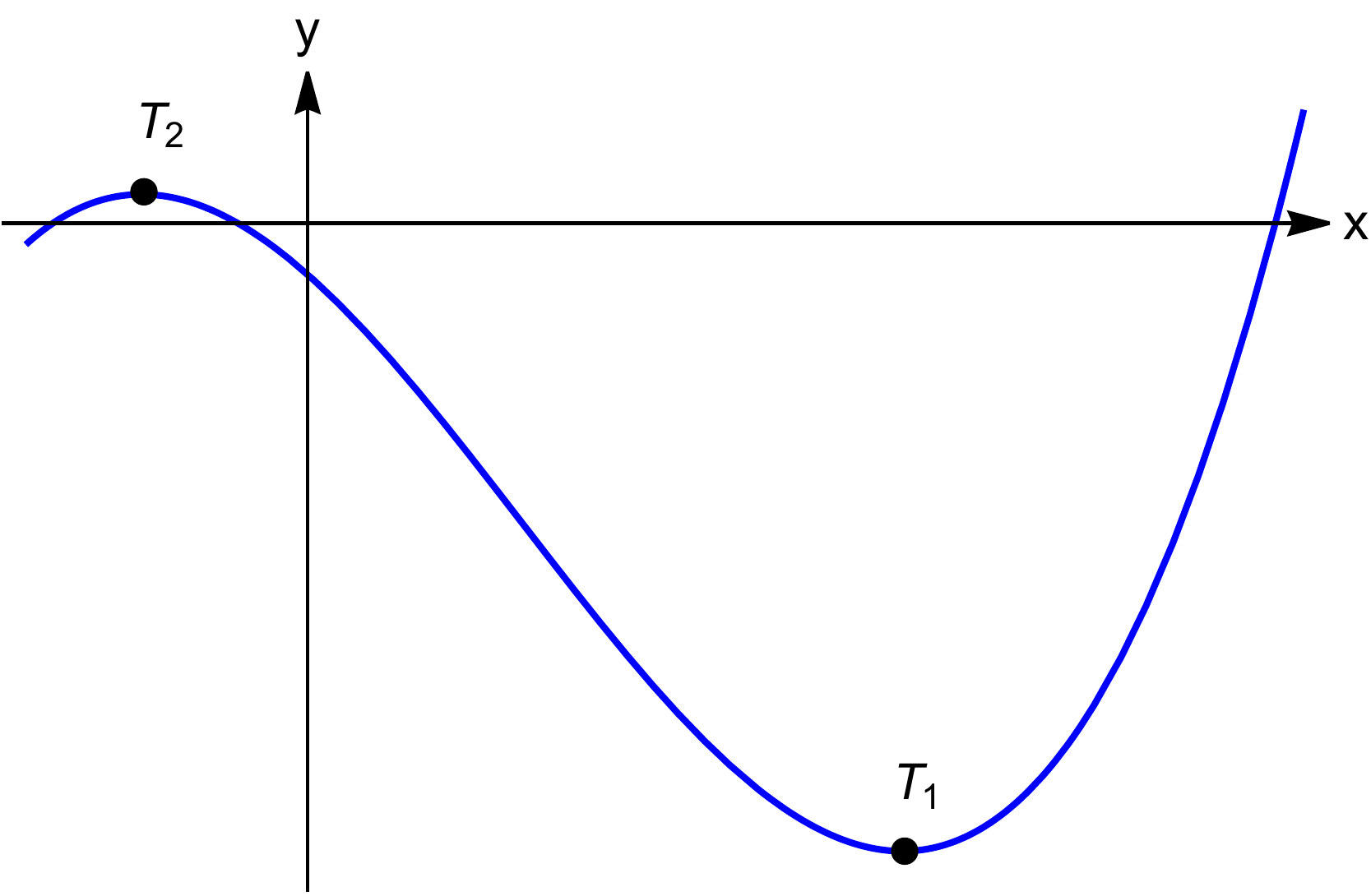}
\caption{Typical plot for section 5.3 with $J\neq 0$ and $\mathcal{A}>0$.}
\label{fig:cas3}
\end{wrapfigure}In this case, $T_{1}$ still belongs to the fourth quadrant,
and $\Delta >0$ implies the existence of three distinct real roots. As $%
\mathcal{A}>0\Rightarrow J^{2}<S_{0}^{2}\,\mathbf{g}$, $T_{2}$ belongs to
the second quadrant and this case is under-rotating, too. Again, since $%
f(0)=-J^{2}S_{0}^{4}$ and observing that $\mathcal{S}_{1}>0$ and $\mathcal{S}%
_{2}<0$, one can realize that only one of the real roots lies on the
positive $x$ axis, and is therefore physically acceptable. A prototypical
plot for this case is shown in Fig. 3 
with $J=2$ and $S_{0}=7$.

\subsection{$J\neq 0$, $\mathcal{A}<0$}

As evident from the previous analysis, $\mathcal{A}<0\Leftrightarrow
J^{2}>S_{0}^{2}\,\mathbf{g}$. In this case, $\Delta <0$, hence there exists
a single real root. The turning point $T_{1}$ is still located in the fourth
quadrant, whereas $T_{2}$ can be either in the third or fourth quadrant.
This situation is rather delicate, as it covers both the under- and over-
rotating class of rotating, stationary, extremal BHs. Depending on the
various possible values of $J^{2}$ there are several situations as shown in
Fig. 4(a) and Fig. 4(b). However, although the location of $T_{1}$ and $%
T_{2} $ changes, the curve $y=f(x)$ (\ref{eq:fcurve}) always looks the same,
and furthermore, it implies the existence of a unique strictly positive real
root of the equation $f(x)=0$, i.e. of the inhomogeneous cubic equation (\ref%
{eq:cubic1}), corresponding to the physically sensible solution.

\begin{figure}[tbph]
\begin{subfigure}{0.48\textwidth}
		\centering
		\includegraphics[width=\textwidth]{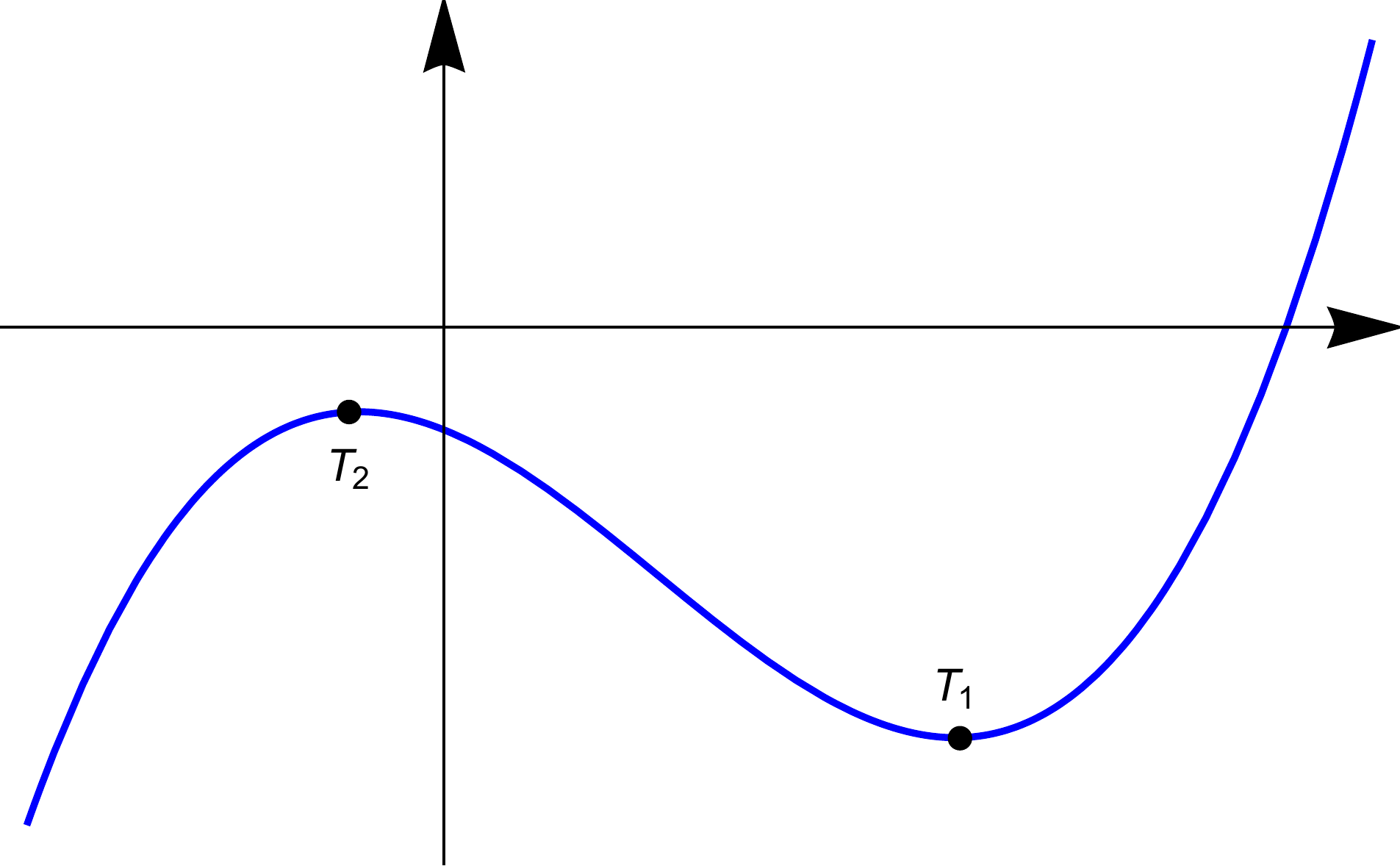}
		\caption{Under-rotating as $S_0^2\mathbf{g}<J^2< {1\over 2}S_0^2$}
  \label{subfig:one}
	\end{subfigure}\hfill
\begin{subfigure}{0.48\textwidth}
		\centering
    \includegraphics[width=\textwidth]{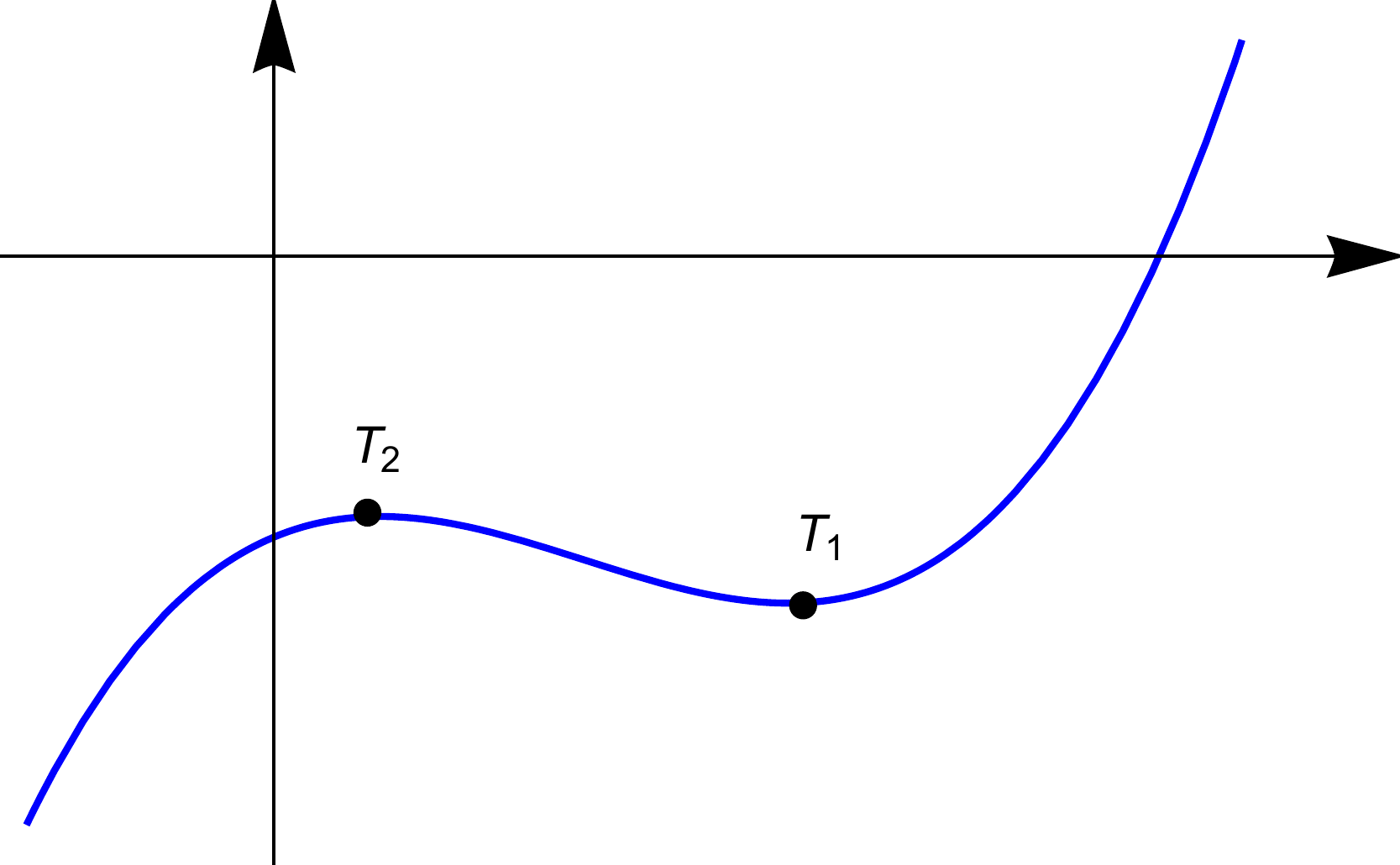}
		\caption{Over(under)-rotating as ${1\over 2}S_0^2<J^2< 2S_0^2$}
  \label{subfig:two}
	\end{subfigure}
\caption{Typical plots for section 5.4 with $S_{0}^{2}\mathbf{g}%
<J^{2}<2S_{0}^{2} $.}
\label{fig:cas4_1}
\end{figure}

\subsubsection{Under-rotating}

Since $J^{2}<S_{0}^{2}\Rightarrow S_{0}^{2}\mathbf{g}<J^{2}<S_{0}^{2}$, both
cases depicted in Fig. 4(a) 
and Fig. 4(b) 
are possible, again with $\mathcal{S}_{1}>0$ and $\mathcal{S}_{2}<0$. The
existence of a unique and physically sensible root is evident.

\subsubsection{Over-rotating}

Besides the case depicted in Fig. 4(b) 
with $S_{0}^{2}<J^{2}<2S_{0}^{2}$, when $J^{2}\geqslant 2S_{0}^{2}$ there
exist two other interesting possibilities :
\begin{figure}[tbph]
\begin{subfigure}{0.48\textwidth}
		\centering
		\includegraphics[width=\textwidth]{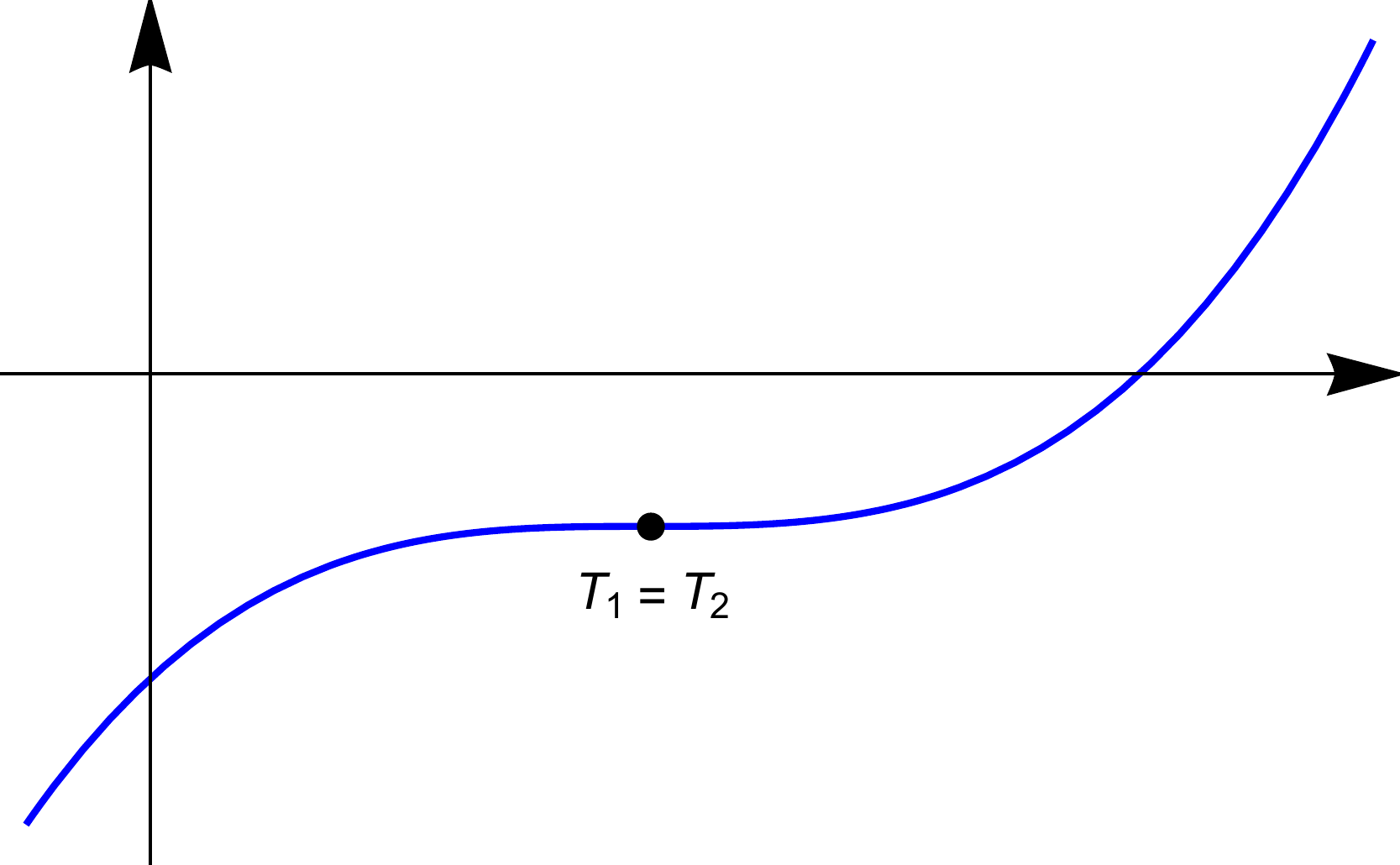}
		\caption{Over-rotating with $J^2=2S_0^2$}
  \label{subfig:three}
	\end{subfigure}\hfill
\begin{subfigure}{0.48\textwidth}
		\centering
		\includegraphics[width=\textwidth]{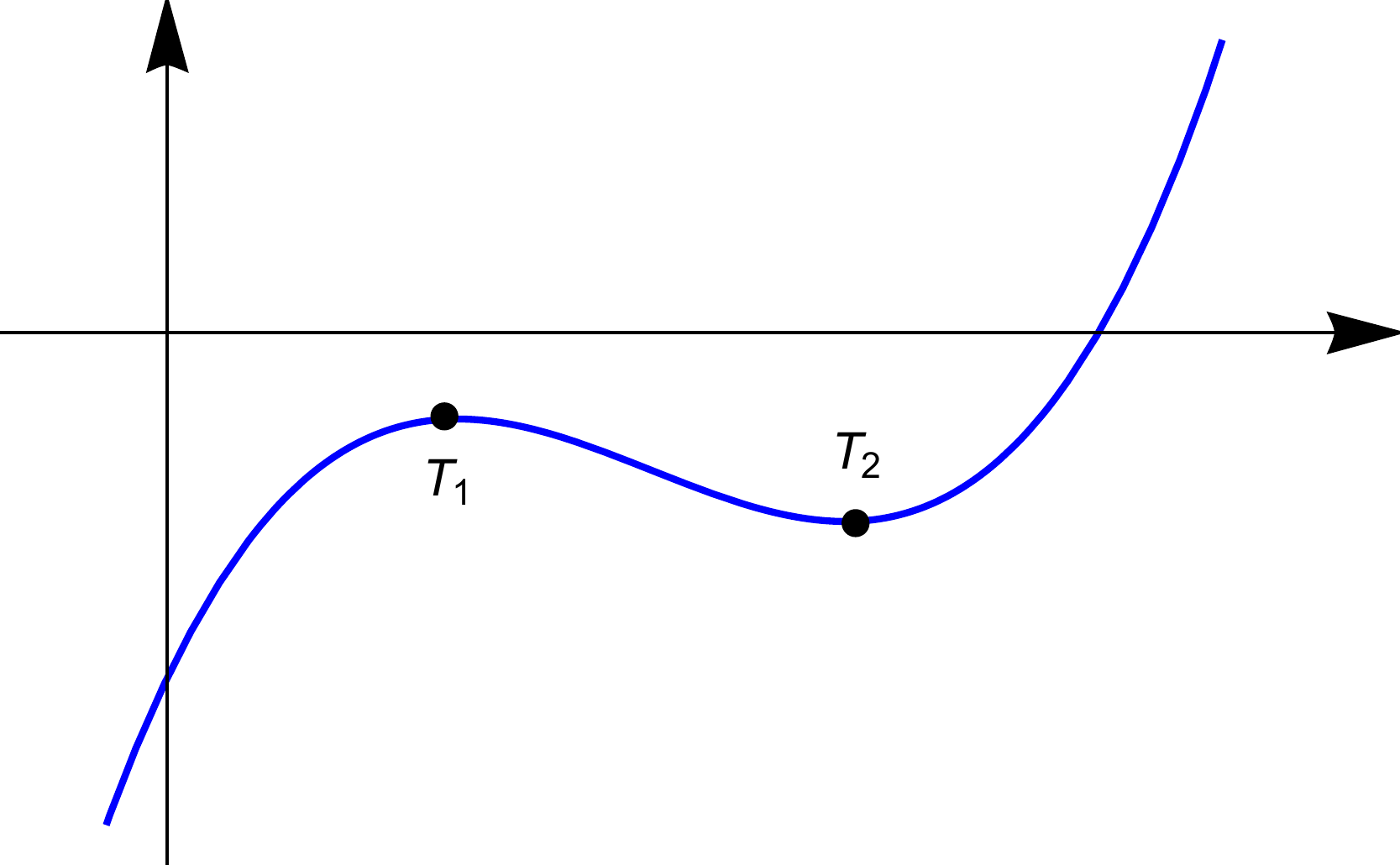}
		\caption{Over-rotating with $J^2>2S_0^2$.}
  \label{subfig:four}
	\end{subfigure}
\caption{Typical plots for section 5.4 with $J^2\geq 2S_0^2$.}
\label{fig:cas4_2}
\end{figure}

\begin{enumerate}
\item The case $J^{2}=2S_{0}^{2}$, which has been already considered in Sec. %
\ref{sec:over}, has $T_{1}=T_{2}$ and $\mathcal{S}_{1}=\mathcal{S}_{2}=0 $,
implying that $y=2S_{0}^{2}$ is the unique (strictly) positive real root.
This (over-rotating) case trivializes the RFD map, which coincides with the
identity, since it leaves both $J$ and $S_{0}$ (and thus $S_{\text{over}}$ (%
\ref{enover})) invariant.

\item The case $J^{2}>2S_{0}^{2}$ has, interestingly, $\mathcal{S}_{1}<0$
and $\mathcal{S}_{2}>0$. However, since both $T_{1}$ and $T_{2}$ swap their
order along the $x$ axis, the equation $f(x)=0$ ends up having a unique
(strictly positive, and thus) physically sensible solution.\medskip
\end{enumerate}

\subsection*{To recap}

The above-detailed analysis proves that, for any value of $J$ and $S_{0}$
physically allowed for stationary (Kerr-Newman, asymptotically flat)
extremal BHs (i.e., $J\in \mathbb{R}^{+}$ and $S_{0}\in \mathbb{R}_{0}^{+}$%
), the (generalized) rotating Freudenthal duality (RFD) map defined in (\ref%
{RFD}) allows for a \textit{unique} RFD-dual BH. The above analysis is
pictorially summarized in Fig. 6, 
in which we plotted all the real roots of \eqref{eq:cubic1} by varying $J\in
\mathbb{R}^{+}$ (for $S_{0}=2$).
\begin{figure}[tbph]
\centering
\includegraphics[width=0.6\textwidth]{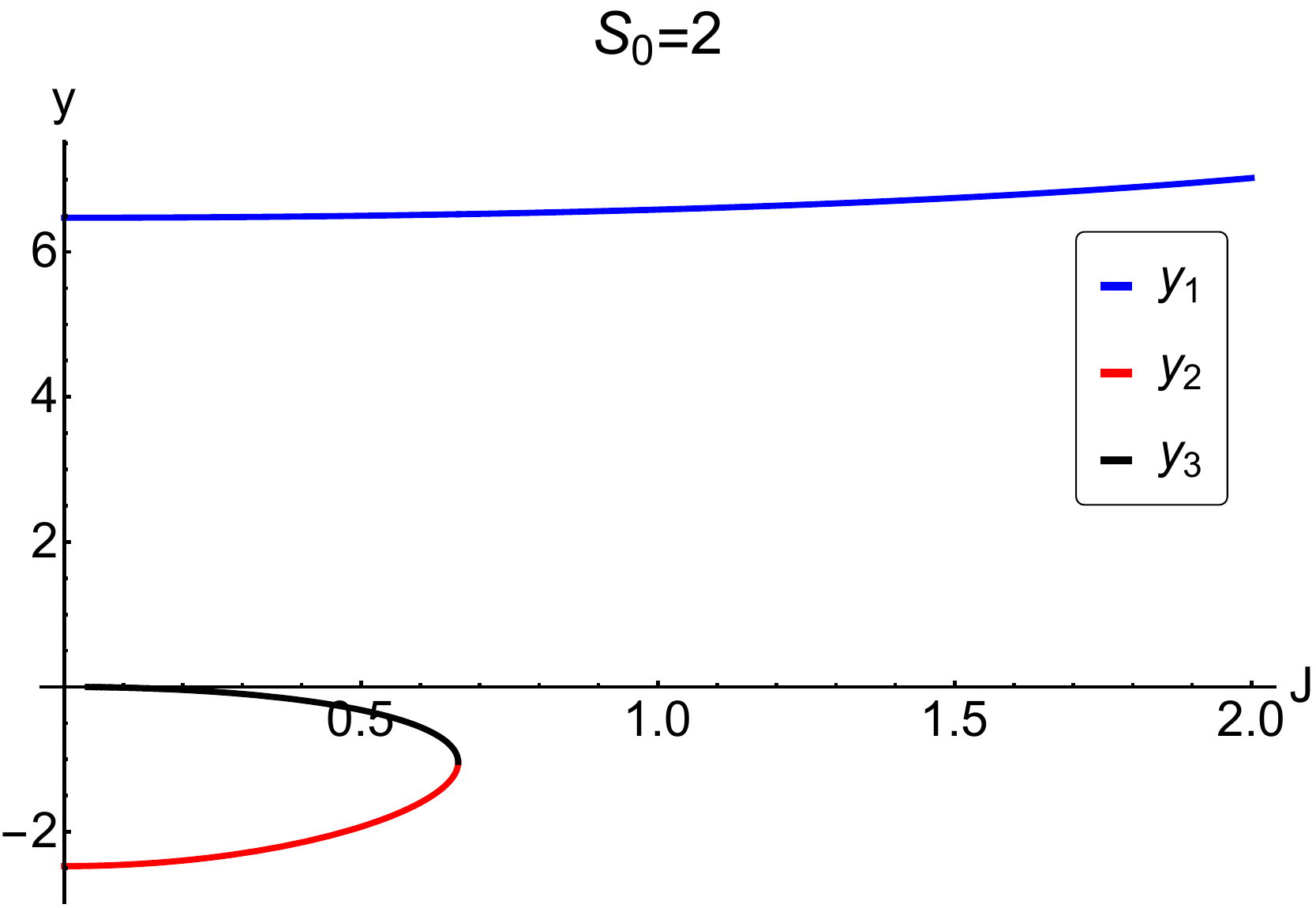}
\caption{Real roots $y_{1}(J)$, $y_{2}(J)$ and $y_{3}(J)$ of the
inhomogeneous cubic equation (\protect\ref{eq:cubic1}) (for $S_{0}=2$).}
\label{fig:allroots}
\end{figure}

\section{\label{sec:an}Analytic form of RFD}

After the qualitative analysis of the previous Section, in which we have
proved that RFD (\ref{RFD}) defines a one-to-one map between two rotating
(stationary, Kerr-Newman, asymptotically flat) extremal BHs with the same
Bekenstein-Hawking entropy, we now discuss the analytical form of the RFD
itself, namely the explicit, analytical form of the solutions to the
inhomogeneous cubic equation (\ref{eq:cubic1}), i.e. of $f(x)=0$, where the
curve $y=f(x)$ is defined in (\ref{eq:fcurve}).

From the general theory of cubic equations, the \textit{depressed form} of
the equation \eqref{eq:cubic1} reads
\begin{equation}
t^{3}+p\,t+q=0,~\text{with}\left\{
\begin{array}{l}
t:=x-\frac{1}{3}{\left( J^{2}+S_{0}^{2}\right) ;} \\
\\
p:=-{\frac{1}{3}}(J^{2}-2S_{0}^{2})^{2}; \\
\\
q:=\frac{1}{54}{\mathcal{A}}-\frac{1}{2}{S_{0}^{6},}%
\end{array}%
\right.  \label{eq:deprsd}
\end{equation}%
and the discriminant can be computed to be (cf. (\ref{discriminant}))
\begin{equation}
\Delta =-(4p^{3}+27q^{2})=S_{0}^{6}\mathcal{A}.
\end{equation}%
We have the following case study.

\subsection{$\Delta <0$}

In this case, $\mathcal{A}<0\Leftrightarrow J^{2}>S_{0}^{2}\,\mathbf{g}$.
Since $\mathbf{g}<1$, the extremal BH can be under-rotating or
over-rotating, depending on whether $S_{0}^{2}\,\mathbf{g}<J^{2}<S_{0}^{2}\,$%
\ or $J^{2}>S_{0}^{2}$, respectively. Moreover, $q<0$, and the equation %
\eqref{eq:cubic1} has one real root and two non-real complex conjugate
roots. By further defining
\begin{eqnarray}
\mathcal{C} &:=&{\frac{q^{2}}{4}}+{\frac{p^{3}}{27}}=-{\frac{S_{0}^{6}}{108}}%
\mathcal{A}=-\frac{\Delta }{108}>0; \\
u &:=&-{\frac{q}{2}}+\sqrt{\mathcal{C}}>0; \\
v &=&-{\frac{q}{2}}-\sqrt{\mathcal{C}}>0,
\end{eqnarray}%
by Cardano's method, the unique real root reads
\begin{equation}
t_{1}=\sqrt[3]{u}+\sqrt[3]{v}>0.  \label{eq:theroot}
\end{equation}%
For completeness, we mention that the other two non-real complex conjugate
roots of \eqref{eq:deprsd} can be written as
\begin{equation}
\begin{split}
t_{2}& =\omega \sqrt[3]{u}+\bar{\omega}\sqrt[3]{v}; \\
& \\
t_{3}& =\bar{\omega}\sqrt[3]{u}+\omega \sqrt[3]{v},
\end{split}
\label{eq:theroott}
\end{equation}%
with $\omega $ and $\bar{\omega}$ being the non-real cube roots of unity,
\begin{equation}
\omega :={\frac{-1+i\,\sqrt{3}}{2}.}
\end{equation}

Thus, by recalling that $x:=\hat{J}^{2}$, the unique physically sensible
solution for the RFD-transformed (\textit{aka} RFD-dual) angular momentum
reads%
\begin{gather}
\hat{J}^{2}:=\left( J+\delta J\right) ^{2}=t_{1}+\frac{1}{3}\left( {%
J^{2}+S_{0}^{2}}\right) >0;  \notag \\
\Updownarrow  \notag \\
\delta J=-J+\sqrt{t_{1}+\frac{1}{3}\left( {J^{2}+S_{0}^{2}}\right) }\in
\mathbb{R},  \label{thiss}
\end{gather}%
with $t_{1}$ given by (\ref{eq:theroot}).

\subsection{$\Delta =0$ (under-rotating)}

In this case, $\mathcal{A}=0\Leftrightarrow J^{2}=S_{0}^{2}\mathbf{g}%
<S_{0}^{2}$ (and thus the extremal BH is under-rotating). Then, since its
coefficients are all real, the equation \eqref{eq:cubic1} has three real
roots, with non-trivial multiplicity. Since $q<0$ and $p<0$, in this case
the roots indeed read%
\begin{eqnarray}
t_{1} &=&3\frac{q}{p}>0;  \label{eq:theroot2} \\
t_{2} &=&t_{3}=-3\frac{q}{p}<0.
\end{eqnarray}
Therefore, the unique physically sensible solution for the RFD-dual angular
momentum still formally reads%
\begin{gather}
\hat{J}^{2}:=t_{1}+\frac{1}{3}\left( {J^{2}+S_{0}^{2}}\right) >0;  \notag \\
\Updownarrow  \notag \\
\delta J=-J+\sqrt{t_{1}+\frac{1}{3}\left( {J^{2}+S_{0}^{2}}\right) }\in
\mathbb{R},
\end{gather}%
but with $t_{1}$ given by (\ref{eq:theroot2}).

\subsection{$\Delta >0$ (under-rotating)}

This particular situation goes under the name of \textit{casus irreducibilis}%
, since at Cardano's time the complex roots were not known to exist. In this
case, $\mathcal{A}>0\Rightarrow J^{2}<S_{0}^{2}\,\mathbf{g}$, and therefore
the extremal BH is under-rotating; from the general theory, there are three
distinct real roots, but, as discussed in Sec. \ref{sec:this}, only one of
them is positive. Remarkably, the same three roots $t_{1}$, $t_{2}$ and $%
t_{3}$ given by Cardano's method (i.e., (\ref{eq:theroot}) and (\ref%
{eq:theroott})) work, \textit{provided one takes the principal value} for
the cube roots in (\ref{eq:theroot}) and (\ref{eq:theroott}), as well as for
the square root in (\ref{thiss}). In fact, in this case $t_{1}$, given by (%
\ref{eq:theroot}), still is the unique physically consistent solution, which
can now more conveniently be written as
\begin{eqnarray}
t_{1} &=&{\frac{1}{3\cdot 2^{2/3}}}\left( 3^{3/2}S_{0}^{3}+\sqrt{-\mathcal{A}%
}\right) ^{2/3}+{\frac{1}{3\cdot 2^{2/3}}}\left( 3^{3/2}S_{0}^{3}-\sqrt{-%
\mathcal{A}}\right) ^{2/3}  \notag \\
&=&{\frac{1}{3\cdot 2^{2/3}}}\left( z^{2/3}+\bar{z}^{2/3}\right) >0;
\label{eq:t1def} \\
\quad z &:=&3^{3/2}S_{0}^{3}+i\,\sqrt{\mathcal{A}}.
\end{eqnarray}%
Since we are taking the principal value here, $t_{1}$ is real and positive,
as the imaginary parts would cancel out in (\ref{eq:t1def}).\medskip

\subsection*{To recap}

Therefore, in all cases, there exists a unique physically sensible root of %
\eqref{eq:cubic1}, which can be cast in the same form, namely%
\begin{eqnarray}
\hat{J}^{2} &:=&t_{1}+\frac{1}{3}\left( {J^{2}+S_{0}^{2}}\right) >0,
\label{eq:finalGFD} \\
\text{or~}\delta J &=&-J+\sqrt{t_{1}+\frac{1}{3}\left( {J^{2}+S_{0}^{2}}%
\right) }\in \mathbb{R},
\end{eqnarray}%
with $t_{1}$ which can generally be written as follows :
\begin{equation}
t_{1}=\left( {\frac{S_{0}^{3}}{2}}+\sqrt{-{\frac{\mathcal{A}}{108}}}\right)
^{\frac{2}{3}}+\left( {\frac{S_{0}^{3}}{2}}-\sqrt{-{\frac{\mathcal{A}}{108}}}%
\right) ^{\frac{2}{3}}.  \label{t1}
\end{equation}

\section{\label{sec:underover}Under- $\rightleftarrows $ Over- rotating
transitions through RFD? \textit{No-Go}}

In the previous Sections, we have proved the uniqueness and presented the
explicit analytic form, of the RFD map introduced in (\ref{RFD}). One might
wonder whether RFD may be responsible for a transition from the
under-rotating regime to the over-rotating one, or \textit{vice versa}. In
this Section, we will answer such a question.

We start by recalling (\ref{enunder}), (\ref{enover}) and (\ref{Q-transf}),
implying that, under RFD (\ref{RFD}), the overall entropy \textit{formally}%
\footnote{%
We stress that the transformation (\ref{ttransf}) is only formal, because (%
\ref{enequal}) crucially holds: the overall Bekenstein-Hawking entropy is
\textit{invariant} under RFD.} transforms as follows :%
\begin{equation}
S_{\text{under(over)}}\left( Q,J\right) \mapsto S_{\text{under(over)}}\left(
\hat{Q}(Q,J+\delta J),J+\delta J\right) ,  \label{ttransf}
\end{equation}%
where (cf. (\ref{enunder}) and (\ref{enover}))%
\begin{equation}
S_{\text{under(over)}}\left( Q,J\right) :=\sqrt{\left\vert
S_{0}^{2}(Q)-J^{2}\right\vert },
\end{equation}%
and%
\begin{eqnarray}
&&S_{\text{under(over)}}\left( \hat{Q}(Q,J+\delta J),J+\delta J\right)
\notag \\
&&= \sqrt{\left\vert S_{0}^{2}\left( \hat{Q}(Q,J+\delta J)\right) -\left(
J+\delta J\right) ^{2}\right\vert }  \notag \\
&&=\sqrt{\left\vert S_{0}^{2}\left( \pm \frac{S_{0}\left( Q\right) }{\sqrt{%
\left\vert S_{0}^{2}\left( Q\right) -\left( J+\delta J\right)
^{2}\right\vert }}\Omega \frac{\partial S_{0}\left( Q\right) }{\partial Q}%
\right) -\left( J+\delta J\right) ^{2}\right\vert }  \notag \\
&&=\sqrt{\left\vert \frac{S_{0}^{4}\left( Q\right) }{\left[ S_{0}^{2}\left(
Q\right) -\left( J+\delta J\right) ^{2}\right] ^{2}}S_{0}^{2}\left( \Omega
\frac{\partial S_{0}\left( Q\right) }{\partial Q}\right) -\left( J+\delta
J\right) ^{2}\right\vert }  \notag \\
&&=\sqrt{\left\vert \frac{S_{0}^{6}\left( Q\right) }{\left[ S_{0}^{2}\left(
Q\right) -\left( J+\delta J\right) ^{2}\right] ^{2}}-\left( J+\delta
J\right) ^{2}\right\vert }  \notag \\
&&=\sqrt{\left\vert \tilde{S_{0}}^{2}\left( Q,\hat{J}\right) -\hat{J}%
^{2}\right\vert }  \label{thisss}
\end{eqnarray}%
Where we have used the properties (\ref{hom-2-Q}) and (\ref{FD-inv}) of $%
S_{0}$ to obtain (\ref{thisss}). Moreover, by recalling that $\hat{J}%
:=J+\delta J$, we have introduced
\begin{equation}
\tilde{S_{0}}\left( Q,\hat{J}\right) \equiv \tilde{S_{0}}\left( Q,J,\delta
J\right) :=\frac{S_{0}^{3}\left( Q\right) }{\left\vert S_{0}^{2}\left(
Q\right) -\hat{J}^{2}\right\vert }\in \mathbb{R}_{0}^{+}.  \label{S-tilde}
\end{equation}

Thus, depending on the sign of $\tilde{S_{0}}^{2}\left( Q,\hat{J}\right) -%
\hat{J}^{2}$, one can establish whether the \textit{formal} transformation (%
\ref{ttransf})-(\ref{thisss}) of the overall Bekenstein-Hawking BH entropy
pertains to an under- or over- rotating extremal BH, i.e.%
\begin{eqnarray}
\tilde{S_{0}}^{2}\left( Q,\hat{J}\right) -\hat{J}^{2} &>&0\Rightarrow \text{%
the~RFD-dual~BH~is~under-rotating}; \\
\tilde{S_{0}}^{2}\left( Q,\hat{J}\right) -\hat{J}^{2} &<&0\Rightarrow \text{%
the~RFD-dual~BH~is~over-rotating}.
\end{eqnarray}
In order to study this sign problem, we introduce the real, non-negative,
dimensionless parameters%
\begin{eqnarray}
\alpha \left( Q,J\right) &:=&\frac{J^{2}}{S_{0}^{2}\left( Q\right) }\in
\mathbb{R}^{+}:\left\{
\begin{array}{l}
0\leqslant \alpha <1:~\text{the~starting~BH~is~under-rotating}; \\
\alpha >1:\text{the~starting~BH~is~over-rotating}.%
\end{array}%
\right.  \label{alpha} \\
&&  \notag \\
\hat{\alpha}\left( Q,\hat{J}\right) &\equiv &\hat{\alpha}\left( Q,J,\delta
J\right) :=\frac{\hat{J}^{2}}{\tilde{S_{0}}^{2}\left( Q,\hat{J}\right) }\in
\mathbb{R}^{+}:\left\{
\begin{array}{l}
0\leqslant \hat{\alpha}<1:~\text{the~RFD-dual~BH~is} \\
\text{~under-rotating}; \\
\hat{\alpha}>1:\text{the~RFD-dual~BH~is} \\
\text{~over-rotating}.%
\end{array}%
\right.  \label{alpha-hat}
\end{eqnarray}%
Such parameters are connected by the RFD map (\ref{RFD}),%
\begin{equation}
\alpha \left( Q,J\right) \overset{\text{RFD~(\ref{RFD})}}{\longrightarrow }%
\hat{\alpha}\left( Q,\hat{J}\right) .
\end{equation}

There are two limit cases :

\begin{enumerate}
\item $\alpha =0\Leftrightarrow J=0$ (non-rotating limit) : the starting
extremal BH is non-rotating (i.e., static); as discussed in Sec. \ref%
{sec:golden}, the RFD map (\ref{RFD}) can yield to $\hat{\alpha}=0$ :%
\begin{equation}
\alpha \left( Q,J=0\right) =0\overset{\text{RFD}_{J\rightarrow 0^{+}}\text{~(%
\ref{RFD})}}{\longrightarrow }\hat{\alpha}\left( Q,\hat{J}=0\right) =0,
\end{equation}
but it also admits another spurious, \textquotedblleft
golden\textquotedblright\ branch, which maps $\alpha $ to (cf. (\ref%
{eq:delJat0}), (\ref{cruc}), (\ref{S-tilde}) and (\ref{alpha-hat})) :%
\begin{eqnarray}
\alpha \left( Q,J=0\right) &=&0\overset{\text{RFD}_{J\rightarrow 0^{+}\text{%
,~golden}}\text{~(\ref{RFD})}}{\longrightarrow }\hat{\alpha}_{\text{golden}},
\\
\text{where~}\hat{\alpha}_{\text{golden}} &:=&\hat{\alpha}\left( Q,\hat{J}_{%
\text{golden}}\right) =\frac{\hat{J}_{\text{golden}}^{2}}{\tilde{S_{0}}%
^{2}\left( Q,\hat{J}_{\text{golden}}\right) }=\phi \left( 1-\phi \right) ^{2}
\notag \\
&&=\frac{1}{\phi }=\phi -1<1,
\end{eqnarray}%
implying that the image of the non-rotating (static) extremal BH under the
\textquotedblleft golden\textquotedblright\ branch of RFD$_{J\rightarrow
0^{+}}$, defined as RFD$_{J\rightarrow 0^{+}\text{,~golden}}$ (\ref{RFD-J=0}%
) in Sec. \ref{sec:golden}, is an \textit{under-rotating} (stationary)
extremal BH; see the discussion, and in particular the Remark 1, in Sec. \ref%
{sec:golden}.

\item $\alpha =1\Leftrightarrow J=S_{0}(Q)$ : from (\ref{enunder}) and (\ref%
{enover}), we find $S_{\text{under}}\left( Q,J=S_{0}\right) =0=S_{\text{over}%
}\left( Q,J=S_{0}\right) $). Hence, the starting BH is \textquotedblleft
small\textquotedblright , i.e. it has a vanishing Bekenstein-Hawking
entropy/area of the (unique) event horizon, \textit{at least} within the
two-derivative (Einsteinian) treatment understood in the present treatment.
The RFD map (\ref{RFD}) yields to (cf. (\ref{S-tilde}) and (\ref{alpha-hat}))%
\begin{eqnarray}
\alpha \left( Q,J=S_{0}\right)  &=&1\overset{\text{RFD~(\ref{RFD})}}{%
\longrightarrow }\check{\alpha}\left( \frac{\delta J}{J}\right) , \\
\text{where~}\check{\alpha}\left( \frac{\delta J}{J}\right)  &:=&\hat{\alpha}%
\left( Q,\hat{J}=S_{0}+\delta J\right)   \notag \\
&=&\frac{\left[ S_{0}\left( Q\right) +\delta J\right] ^{2}\left[
S_{0}^{2}\left( Q\right) -\left( S_{0}\left( Q\right) +\delta J\right) ^{2}%
\right] ^{2}}{S_{0}^{6}(Q)}  \notag \\
&=&\frac{\left( \delta J\right) ^{2}\left[ S_{0}\left( Q\right) +\delta J%
\right] ^{2}\left[ 2S_{0}(Q)+\delta J\right] ^{2}}{S_{0}^{6}(Q)}  \notag \\
&=&\left( \frac{\delta J}{J}\right) ^{2}\left( 1+\frac{\delta J}{J}\right)
^{2}\left( 2+\frac{\delta J}{J}\right) ^{2}.  \label{alpha-hhat}
\end{eqnarray}%
However, since the RFD map (\ref{RFD}) preserves (by definition) the overall
Bekenstein-Hawking BH entropy, the RFD-dual extremal BH is also
\textquotedblleft small\textquotedblright , with%
\begin{equation}
S_{\text{under}}\left( Q,J=S_{0}\right) =0=S_{\text{over}}\left(
Q,J=S_{0}\right) \mapsto S\left( \hat{Q}(J+\delta J),J+\delta J\right) =0,
\end{equation}%
implying that (cf. (\ref{alpha-hhat}))%
\begin{gather}
\check{\alpha}\left( \frac{\delta J}{J}\right) =1 \\
\Updownarrow   \notag \\
\left( \frac{\delta J}{J}\right) ^{2}\left( 1+\frac{\delta J}{J}\right)
^{2}\left( 2+\frac{\delta J}{J}\right) ^{2}=1 \\
\Updownarrow   \notag \\
\left( \delta J\right) ^{2}\left( J+\delta J\right) ^{2}\left( 2J+\delta
J\right) ^{2}=J^{6},
\end{gather}%
This is an inhomogeneous algebraic equation of degree $6$ in $\delta J$,
equivalently obtained by plugging $S_{0}=J$ in (\ref{eq:sextic1}) and (\ref%
{cond-6}) :
\begin{eqnarray}
0 &=&a_{1}\left( \delta J\right) ^{6}+a_{2}\left( \delta J\right)
^{5}+a_{3}\left( \delta J\right) ^{4}+a_{4}\left( \delta J\right)
^{3}+a_{5}\left( \delta J\right) ^{2}+a_{6}\delta J+a_{7},  \label{eq:jeqs0}
\\
a_{1} &=&1;  \notag \\
a_{2} &=&6J;  \notag \\
a_{3} &=&13J^{2};  \notag \\
a_{4} &=&12J^{3};  \notag \\
a_{5} &=&4J^{4};  \notag \\
a_{6} &=&0;  \notag \\
a_{7} &=&-J^{6}.
\end{eqnarray}%
Solving \eqref{eq:jeqs0}, the unique real and positive (and thus physically
meaningful) solution reads $\delta J\simeq 0.324718J$; the details are
provided in App. \ref{app:solj}\footnote{%
With \eqref{falpha} and appendix \ref{app:solj} one can check the relation $%
\left( {\frac{\hat{J}}{J}}\right) ^{2}=f(1)$.}. However, since both the
starting and the RFD-dual extremal BHs are \textquotedblleft
small\textquotedblright , this limit case is not interesting, \textit{at
least} at two-derivative level.\bigskip
\end{enumerate}

Reconsidering the general treatment, by using the explicit form of the
physically sensible root of (\ref{eq:sextic1}), namely (\ref{eq:finalGFD})
and (\ref{t1}), and recalling the definitions (\ref{eq:Aeq}) of $\mathcal{A}$%
, (\ref{alpha}) of $\alpha $, and (\ref{S-tilde}) of $\tilde{S}_{0}$, one
can compute\footnote{$S_{0}\equiv S_{0}(Q)$ throughout.}%
\begin{eqnarray}
\hat{J}^{2} &=&t_{1}+\frac{1}{3}\left( {J^{2}+S_{0}^{2}}\right)  \notag \\
&=&\left( {\frac{S_{0}^{3}}{2}}+\sqrt{-{\frac{\mathcal{A}}{108}}}\right) ^{%
\frac{2}{3}}+\left( {\frac{S_{0}^{3}}{2}}-\sqrt{-{\frac{\mathcal{A}}{108}}}%
\right) ^{\frac{2}{3}}+\frac{1}{3}\left( {J^{2}+S_{0}^{2}}\right)  \notag \\
&=&\frac{S_{0}^2}{2^{2/3}}\left(\left( {1}+\sqrt{1+\frac{{4}}{27}{(\alpha
-2)^{3}}}\right) ^{\frac{2}{3}}+\left( {1}-\sqrt{1+\frac{{4}}{27}{(\alpha
-2)^{3}}}\right) ^{\frac{2}{3}}\right)+\frac{S_{0}^2}{3}\left( {\alpha +1}%
\right)  \notag \\
&=&S_{0}^{2}\,f(\alpha ),  \label{eq:JvsS}
\end{eqnarray}%
and%
\begin{equation}
\tilde{S_{0}}^{2}=S_{0}^{2}\left( 1-f(\alpha )\right) ^{-2},
\label{eq:JvsS2}
\end{equation}%
where we have defined the following non-negative function of $\alpha $:%
\begin{equation}
f(\alpha ):={\frac{1}{2^{2/3}}}\left( 1+\sqrt{1+{\frac{4}{27}}(\alpha -2)^{3}%
}\right) ^{2/3}+{\frac{1}{2^{2/3}}}\left( 1-\sqrt{1+{\frac{4}{27}}(\alpha
-2)^{3}}\right) ^{2/3}+{\frac{1}{3}}(\alpha +1),  \label{falpha}
\end{equation}

\begin{figure}[tbph]
\centering
\includegraphics[width=0.6\textwidth]{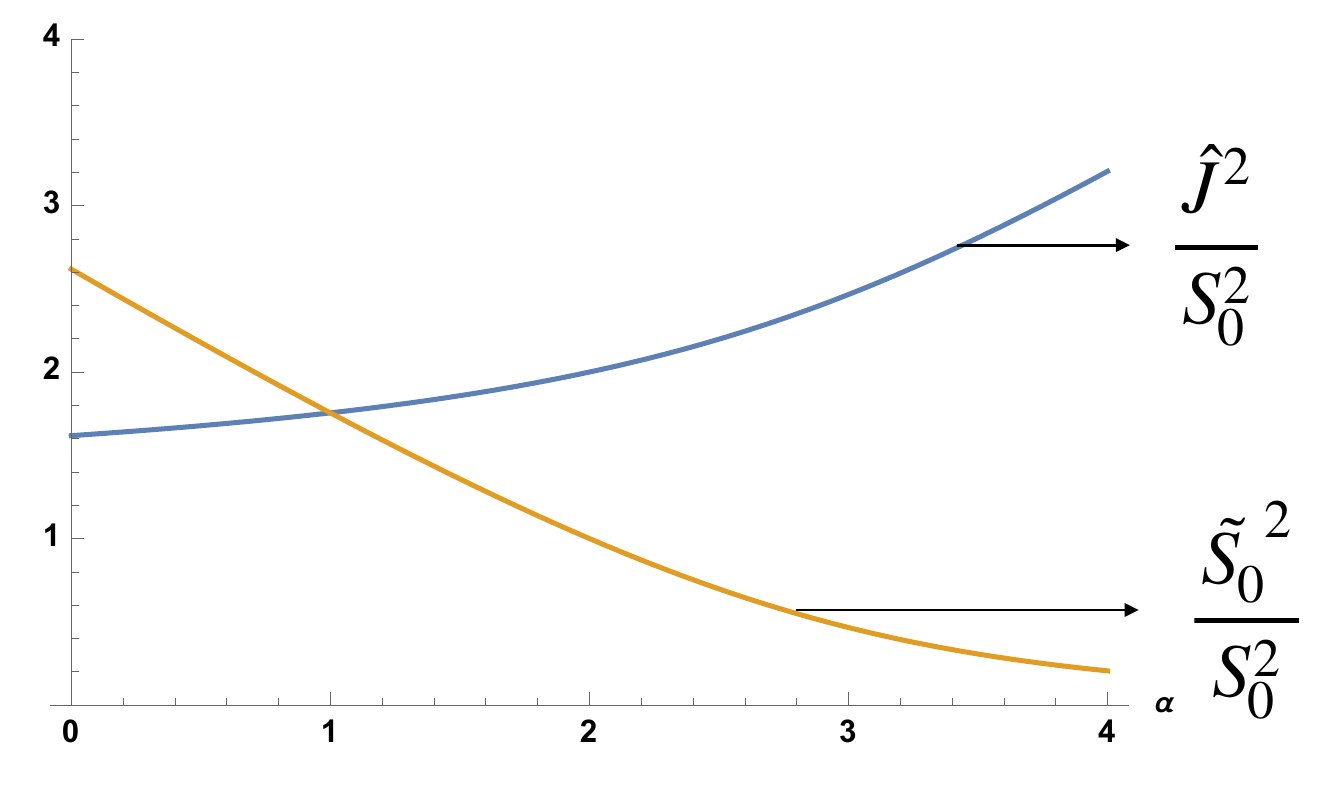
}
\caption{The yellow curve is for $\frac{\tilde{S_{0}}^{2}}{S_{0}^{2}}$ and
the blue one is for $\frac{\hat{J}^{2}}{S_{0}^{2}}$.}
\label{fig:JvsS}
\end{figure}
We can now discuss the possibility of under $\rightleftarrows $ over
-rotating transition by means of the RFD map (\ref{RFD}), by conveniently
plotting $\hat{J}^{2}/S_{0}^{2}=f(\alpha )$ and $\tilde{S}%
_{0}^{2}/S_{0}^{2}=\left( 1-f(\alpha )\right) ^{-2}$ vs. $\alpha $ in Fig.
7. Indeed, we have to study the sign of%
\begin{equation}
\tilde{S}_{0}^{2}-\hat{J}^{2}=\left[ \left( 1-f\left( \alpha \right) \right)
^{-2}-f(\alpha )\right] S_{0}^{2},
\end{equation}%
depending on the value of $\alpha =J^{2}/S_{0}^{2}\geqslant 0$ (by recalling
that for $0\leqslant \alpha <1$ the extremal BH is under-rotating, while for
$\alpha >1$ it is over-rotating). By defining $z:=f\left( \alpha \right) $,
it holds that%
\begin{equation}
\tilde{S}_{0}^{2}-\hat{J}^{2}=0\Leftrightarrow \left( 1-z\right)
^{-2}=z\Leftrightarrow z\left( 1-z\right) ^{2}=1.
\end{equation}%
This peculiar cubic equation has been solved in App. \ref{app:solj} (cf. Eq.
(\ref{solsol})), giving as unique\footnote{%
Under the condition that $f\left( \alpha \right) \neq 1$. This condition is
however always satisfied, since, as we will prove in App. \ref{app:proof}, $%
f(\alpha )$ is monotone increasing for $\alpha \geqslant 0$, and $f\left(
0\right) =4/3$, thus implying that $f(\alpha )\geqslant 4/3>1$ $\forall
\alpha \geqslant 0$.} real non-negative solution\footnote{%
From the treatment given in App. \ref{app:solj}, the value of $f(1)$ can be
computed to be $f(1)=\left. \frac{\hat{J}^{2}}{S_{0}^{2}}\right\vert
_{J=S_{0}}=\frac{\left( S_{0}+\left. \delta J\right\vert _{J=S_{0}}\right)
^{2}}{S_{0}^{2}}\simeq \left( 1.324718\right) ^{2}\simeq 1.754877$.} (cf.
Eq. (\ref{solsol2}))%
\begin{equation}
z=f(1)\Leftrightarrow f\left( \alpha \right) =f(1).\label{this}
\end{equation}%
In App. \ref{app:proof} we prove that $f\left( \alpha \right) $ is monotone
increasing for $\alpha \geqslant 0$; thus, the result (\ref{this}) holds iff
$\alpha =1$, and the following \textit{no-go} result is achieved :%
\begin{eqnarray}
0 &\leqslant &\alpha <1\Rightarrow \tilde{S}_{0}^{2}-\hat{J}^{2}>0; \\
\alpha  &>&1\Rightarrow \tilde{S}_{0}^{2}-\hat{J}^{2}<0.
\end{eqnarray}%
In other words, no under $\rightleftarrows $ over -rotating transition can
be achieved by means of the RFD map (\ref{RFD}), starting from an
under-rotating resp. over-rotating stationary extremal BH.

\section{\label{sec:conc}Conclusion}

In this work, we focused on extremal (stationary, asymptotically flat)
rotating BHs in four space-time dimensions, and we have shown that there
exists a (generally non-anti-involutive) non-linear symmetry of their
Bekenstein-Hawking entropy, both in the under-rotating and in the
over-rotating regime. We have named such a non-linear symmetry (generalized)
rotating Freudenthal duality (RFD): this map generalizes, in the presence of
non-vanishing (constant) angular momentum $J$, the usual Freudenthal duality
(FD) for non-rotating (static) extremal BHs introduced in \cite%
{Ferrara:2011gv}, and it is part of the program, started in \cite%
{Chattopadhyay:2021vog} and \cite{Chattopadhyay:2022ycb} dealing with
near-extremal non-rotating BHs, to extend FD to various classes of BH
solutions beyond the extremal static one. The RFD map has been defined as an
intrinsically non-linear map acting both on the e.m. dyonic charges of the
BH (collectively denoted by the symplectic vector $Q$) and on its angular
momentum $J$, and keeping the Bekenstein-Hawking BH entropy $S$ invariant.
We have proved the uniqueness of the RFD map, and we also explicitly
computed the analytical expression of the transformation of the angular
momentum $J$ under such a map, which generally is a quite involved function
of the charges of the starting extremal BH (through the non-rotating BH
entropy $S_{0}\left( Q\right) $) as well as of its angular momentum $J$
itself.

In the non-rotating limit, RFD reduces to the non-rotating, usual
(anti-involutive) definition of FD (introduced, with the name of
\textquotedblleft on-shell\textquotedblright\ FD, in \cite{Ferrara:2011gv}).
However, there also exists a spurious, \textit{\textquotedblleft
golden\textquotedblright } branch of the RFD map when $J\rightarrow 0^{+}$ ,
which maps a non-rotating (static) extremal BH (with near-horizon geometry $%
AdS_{2}\otimes S^{2}$, and Bekenstain-Hawking entropy $S_{0}(Q)$) to an
under-rotating (stationary) extremal BH (with near-horizon geometry $%
AdS_{2}\otimes S^{1}$), with a (constant) angular momentum given by $\sqrt{%
\phi }S_{0}(Q)$, where $\phi $ is the \textit{golden ratio}; it is then
noteworthy that both such BHs have the same Bekenstein-Hawking entropy!
Furthermore, one can state that the space of (generalized) FD functionals
undergoes a sort of phase transition in the non-rotating limit $J\rightarrow
0^{+}$, such that $J=0$ appears to be a bifurcation point between the
non-rotating usual FD branch and the aforementioned \textit{%
\textquotedblleft golden\textquotedblright } branch. It is suggestive to
remark that the duality between a non-rotating extremal BH and an
under-rotating extremal BH could be traced back to a common parent
five-dimensional extremal BH \cite{LopesCardoso:2007hen,Goldstein:2007km}.
This observation needs further investigation, and we leave that as a future
endeavour.

It will also be interesting to investigate the effect of RFD on the dual CFT$%
_{2}$ of an extremal Kerr-Newman BH, by exploiting the so-called Kerr/CFT
correspondence. On the other hand, we should recall that so far we have only
formulated the generalization of the Freudenthal duality map for the
semi-classical Bekenstein-Hawking entropy of various classes of BHs; the
invariance of the quantum-corrected BH entropy under a suitable
generalization of Freudenthal duality still stands as a crucial issue, which
we hope to address in future investigation.

\section*{\noindent \textbf{Acknowledgments}}

The work of AC is supported by the European Union Horizon 2020 research and
innovation programme under the Marie Sk\l {}odowska Curie grant agreement
number 101034383. The work of TM is supported by the grant
SB/SJF/2019-20/08. The work of AM is supported by a \textquotedblleft Maria
Zambrano\textquotedblright\ distinguished researcher fellowship at the
University of Murcia, Spain, financed by the European Union within the
NextGenerationEU programme.

\appendix

\section{\label{app:solj}Solution for $J=S_{0}$}

We will start by simplifying \eqref{eq:jeqs0} as
\begin{equation}
(\hat{J}^{2}-J^{2})^{2}\hat{J}^{2}=J^{6},
\end{equation}%
with $\hat{J}=J+\delta J$. This can further be simplified as
\begin{equation}
(y-1)^{2}y=1,\label{solsol}
\end{equation}%
by defining $\displaystyle{y={\frac{\hat{J}^{2}}{J^{2}}}}.$ The depressed
form of the equation is then
\begin{equation}
\tilde{t}^{3}-{\frac{\tilde{t}}{3}}-{\frac{25}{27}}=0,  \label{eq:ttilde}
\end{equation}%
with $\tilde{t}=y-{\frac{2}{3}}$. Evidently, \eqref{eq:ttilde} is the same as %
\eqref{eq:deprsd} with the identification that $\displaystyle{\tilde{t}%
\equiv {\frac{t}{J^{2}}}}$ and $p$ and $q$ evaluated at $\alpha =1$.
Therefore we have the relation that
\begin{equation}
{\frac{\hat{J}^{2}}{J^{2}}}={\frac{\hat{J}^{2}}{S_{0}^{2}}}=y=\tilde{t}+{%
\frac{2}{3}}=f(1),\label{solsol2}
\end{equation}%
with $f(\alpha )$ being defined in \eqref{falpha}.

\section{\label{app:proof}Proof of monotonicity of $f(\protect\alpha )$}

We will here prove that both $f(\alpha )$ and $\left( f(\alpha )-1\right)
^{-2}$ are monotone (increasing resp. decreasing) functions $\forall \alpha
\geqslant 0$.

We start and recall Eq. (\ref{eq:deprsd}),%
\begin{equation}
t^{3}+p\,t+q=0,~\text{with~}\left\{
\begin{array}{l}
t:=\hat{J}^{2}-\frac{1}{3}{\left( J^{2}+S_{0}^{2}\right) =}g(\alpha ){S}%
_{0}^{2}{;} \\
\\
p:=-{\frac{1}{3}}(J^{2}-2S_{0}^{2})^{2}=S_{0}^{4}\tilde{p}\left( \alpha
\right) ; \\
\\
q:=-\frac{2}{27}\left( J^{2}-2S_{0}^{2}\right) ^{3}-{S_{0}^{6}=S}_{0}^{6}%
\tilde{q}\left( \alpha \right) {,}%
\end{array}%
\right.
\end{equation}%
where the relations $J^{2}=\alpha S_{0}^{2}$ and $\hat{J}^{2}=f(\alpha
)S_{0}^{2}$ have been used, and the following definitions have been
introduced :%
\begin{eqnarray}
g\left( \alpha \right)  &:=&f\left( \alpha \right) -\frac{\left( \alpha
+1\right) }{3}=\left( \frac{1}{2}+\frac{1}{2}\sqrt{1+{4}\left( \frac{\alpha
-2}{3}\right) ^{3}}\right) ^{2/3}+\left( \frac{1}{2}-\frac{1}{2}\sqrt{1+{4}%
\left( \frac{\alpha -2}{3}\right) ^{3}}\right) ^{2/3};\notag \\ \\
\tilde{p}\left( \alpha \right)  &:=&-\frac{1}{3}\left( \alpha -2\right) ^{2};%
\label{p-tilde} \\ 
\tilde{q}\left( \alpha \right)  &:=&-\frac{1}{27}\left[ 2\left( \alpha
-2\right) ^{3}-1\right] .\label{q-tilde}
\end{eqnarray}%
Therefore, by understanding the dependence on $\alpha $, the depressed cubic
(\ref{eq:deprsd}) can be written as ($S_{0}\equiv S_{0}(Q)>0$ throughout)%
\begin{equation}
g^{3}+\tilde{p}g+\tilde{q}=0,
\end{equation}%
whose first derivative then reads%
\begin{gather}
3g^{2}\frac{dg}{d\alpha }+\frac{d\tilde{p}}{d\alpha }g+\tilde{p}\frac{dg}{%
d\alpha }+\frac{d\tilde{q}}{d\alpha }=0; \\
\Updownarrow   \notag \\
\frac{dg}{d\alpha }=-\frac{\left( \frac{d\tilde{p}}{d\alpha }g+\frac{d\tilde{%
q}}{d\alpha }\right) }{3g^{2}+\tilde{p}}=\frac{2\left( \alpha -2\right) }{%
3\left( 3g-\alpha +2\right) },\label{dg}
\end{gather}%
where the definitions (\ref{p-tilde}) and (\ref{q-tilde}) have been
exploited, respectively implying
\begin{eqnarray}
\frac{d\tilde{p}}{d\alpha } &=&-\frac{2\left( \alpha -2\right) }{3}; \\
\frac{d\tilde{q}}{d\alpha } &=&-\frac{2\left( \alpha -2\right) ^{2}}{9}.
\end{eqnarray}%
Thus, since $f\left( \alpha \right) =g\left( \alpha \right) +\frac{\left(
\alpha +1\right) }{3}$, one obtains%
\begin{equation}
\frac{df}{d\alpha }=\frac{dg}{d\alpha }+\frac{1}{3}=\frac{3g+\alpha -2}{%
3\left( 3g-\alpha +2\right) }=\frac{f-1}{3f-2\alpha +1}.
\end{equation}

On the other hand, by defining%
\begin{equation}
k\left( \alpha \right) :=\left( 1-f\left( \alpha \right) \right) ^{-2},
\end{equation}%
one can compute that%
\begin{equation}
\frac{dk}{d\alpha }=\frac{2}{\left( 1-f\right) ^{3}}\frac{df}{d\alpha }=-%
\frac{2}{\left( f-1\right) ^{2}\left( 3g-\alpha +2\right) }.
\end{equation}

Thus, we obtain that%
\begin{eqnarray}
f &\equiv &f\left( \alpha \right) ~\text{monotone~\textit{increasing}}%
\Leftrightarrow \frac{df}{d\alpha }>0\Leftrightarrow \left\{
\begin{array}{l}
3g+\alpha -2\gtrless 0, \\
3g-\alpha +2\gtrless 0,%
\end{array}%
\right. \forall \alpha \geqslant 0, \\
&&\text{and}  \notag \\
k &\equiv &k\left( \alpha \right) ~\text{monotone~\textit{decreasing}}%
\Leftrightarrow \frac{dk}{d\alpha }<0\Leftrightarrow 3g-\alpha +2>0,~\forall
\alpha \geqslant 0.
\end{eqnarray}%
Thence, in order to prove that both $f(\alpha )$ and $\left( f(\alpha
)-1\right) ^{-2}$ are monotone (increasing resp. decreasing) functions $%
\forall \alpha \geqslant 0$, we have to show that%
\begin{eqnarray}
3g+\alpha -2 &>&0,\forall \alpha \geqslant 0;\label{uno} \\
~3g-\alpha +2 &>&0,~\forall \alpha \geqslant 0.\label{due}
\end{eqnarray}

(\ref{due}) can be proven by observing that, since%
\begin{equation}
\mathcal{G}\left( x;\gamma \right) :=\left( \frac{1}{2}+\frac{1}{2}\sqrt{1+{%
\gamma x}^{3}}\right) ^{2/3}+\left( \frac{1}{2}-\frac{1}{2}\sqrt{1+{\gamma }%
x^{3}}\right) ^{2/3}>x,~\forall \gamma \in \left( 1,\infty \right) ,
\end{equation}
it holds that
\begin{equation}
g\left( \alpha \right) =\mathcal{G}\left( \frac{\alpha -2}{3};4\right) >%
\frac{\alpha -2}{3}\Leftrightarrow 3g-\alpha +2>0,~\forall \alpha \geqslant
0.
\end{equation}

On the other hand, (\ref{dg}) implies that $g\left( \alpha \right) $ is
minimized at $\alpha =2$, at which it acquires the value $g\left( 2\right) =1
$. Thus, (\ref{due}) and (\ref{dg}) imply (\ref{uno}) :%
\begin{equation}
3g+\alpha -2>1>0,\forall \alpha \geqslant 0.~~\square
\end{equation}

\bibliographystyle{jhep}
\bibliography{bib_RFD}

\providecommand{\href}[2]{#2}\begingroup\raggedright\begin{thebibliography}{10}

\bibitem{McClintock:2006xd}
J.~E. McClintock, R.~Shafee, R.~Narayan, R.~A. Remillard, S.~W. Davis and L.-X.
  Li, \emph{{The Spin of the Near-Extreme Kerr Black Hole GRS 1915+105}},
  \href{http://dx.doi.org/10.1086/508457}{\emph{Astrophys. J.} {\bf 652} (2006)
  518--539}, [\href{http://arxiv.org/abs/astro-ph/0606076}{{\tt
  astro-ph/0606076}}].

\bibitem{Borsten:2009zy}
L.~Borsten, D.~Dahanayake, M.~J. Duff and W.~Rubens, \emph{{Black holes
  admitting a Freudenthal dual}},
  \href{http://dx.doi.org/10.1103/PhysRevD.80.026003}{\emph{Phys. Rev. D} {\bf
  80} (2009) 026003}, [\href{http://arxiv.org/abs/0903.5517}{{\tt 0903.5517}}].

\bibitem{Ferrara:2011gv}
S.~Ferrara, A.~Marrani and A.~Yeranyan, \emph{{Freudenthal Duality and
  Generalized Special Geometry}},
  \href{http://dx.doi.org/10.1016/j.physletb.2011.06.031}{\emph{Phys. Lett. B}
  {\bf 701} (2011) 640--645}, [\href{http://arxiv.org/abs/1102.4857}{{\tt
  1102.4857}}].

\bibitem{Borsten:2012pd}
L.~Borsten, M.~J. Duff, S.~Ferrara and A.~Marrani, \emph{{Freudenthal Dual
  Lagrangians}},
  \href{http://dx.doi.org/10.1088/0264-9381/30/23/235003}{\emph{Class. Quant.
  Grav.} {\bf 30} (2013) 235003}, [\href{http://arxiv.org/abs/1212.3254}{{\tt
  1212.3254}}].

\bibitem{Klemm:2017xxk}
D.~Klemm, A.~Marrani, N.~Petri and M.~Rabbiosi, \emph{{Nonlinear symmetries of
  black hole entropy in gauged supergravity}},
  \href{http://dx.doi.org/10.1007/JHEP04(2017)013}{\emph{JHEP} {\bf 04} (2017)
  013}, [\href{http://arxiv.org/abs/1701.08536}{{\tt 1701.08536}}].

\bibitem{Marrani:2012uu}
A.~Marrani, C.-X. Qiu, S.-Y.~D. Shih, A.~Tagliaferro and B.~Zumino,
  \emph{{Freudenthal Gauge Theory}},
  \href{http://dx.doi.org/10.1007/JHEP03(2013)132}{\emph{JHEP} {\bf 03} (2013)
  132}, [\href{http://arxiv.org/abs/1208.0013}{{\tt 1208.0013}}].

\bibitem{Galli:2012ji}
P.~Galli, P.~Meessen and T.~Ortin, \emph{{The Freudenthal gauge symmetry of the
  black holes of N=2,d=4 supergravity}},
  \href{http://dx.doi.org/10.1007/JHEP05(2013)011}{\emph{JHEP} {\bf 05} (2013)
  011}, [\href{http://arxiv.org/abs/1211.7296}{{\tt 1211.7296}}].

\bibitem{Fernandez-Melgarejo:2013ksa}
J.~J. Fernandez-Melgarejo and E.~Torrente-Lujan, \emph{{$N=2$ SUGRA BPS
  Multi-center solutions, quadratic prepotentials and Freudenthal
  transformations}},
  \href{http://dx.doi.org/10.1007/JHEP05(2014)081}{\emph{JHEP} {\bf 05} (2014)
  081}, [\href{http://arxiv.org/abs/1310.4182}{{\tt 1310.4182}}].

\bibitem{Mandal:2017ioi}
A.~Marrani, P.~K. Tripathy and T.~Mandal, \emph{{Supersymmetric Black Holes and
  Freudenthal Duality}},
  \href{http://dx.doi.org/10.1142/S0217751X17501147}{\emph{Int. J. Mod. Phys.
  A} {\bf 32} (2017) 1750114}, [\href{http://arxiv.org/abs/1703.08669}{{\tt
  1703.08669}}].

\bibitem{Borsten:2018djw}
L.~Borsten, M.~J. Duff and A.~Marrani, \emph{{Freudenthal duality and conformal
  isometries of extremal black holes}},
  \href{http://arxiv.org/abs/1812.10076}{{\tt 1812.10076}}.

\bibitem{Borsten:2019xas}
L.~Borsten, M.~J. Duff, J.~J. Fern\'andez-Melgarejo, A.~Marrani and
  E.~Torrente-Lujan, \emph{{Black holes and general Freudenthal
  transformations}},
  \href{http://dx.doi.org/10.1007/JHEP07(2019)070}{\emph{JHEP} {\bf 07} (2019)
  070}, [\href{http://arxiv.org/abs/1905.00038}{{\tt 1905.00038}}].

\bibitem{Cremmer:1978ds}
E.~Cremmer and B.~Julia, \emph{{The N=8 Supergravity Theory. 1. The
  Lagrangian}},
  \href{http://dx.doi.org/10.1016/0370-2693(78)90303-9}{\emph{Phys. Lett. B}
  {\bf 80} (1978) 48}.

\bibitem{Cremmer:1979up}
E.~Cremmer and B.~Julia, \emph{{The SO(8) Supergravity}},
  \href{http://dx.doi.org/10.1016/0550-3213(79)90331-6}{\emph{Nucl. Phys. B}
  {\bf 159} (1979) 141--212}.

\bibitem{Hull:1994ys}
C.~M. Hull and P.~K. Townsend, \emph{{Unity of superstring dualities}},
  \href{http://dx.doi.org/10.1016/0550-3213(94)00559-W}{\emph{Nucl. Phys. B}
  {\bf 438} (1995) 109--137}, [\href{http://arxiv.org/abs/hep-th/9410167}{{\tt
  hep-th/9410167}}].

\bibitem{Marrani:2015wra}
A.~Marrani, \emph{{Freudenthal Duality in Gravity: from Groups of Type E7 to
  Pre-Homogeneous Spaces}},
  \href{http://dx.doi.org/10.1134/S207004661504007X}{\emph{p Adic Ultra. Anal.
  Appl.} {\bf 7} (2015) 322--331}, [\href{http://arxiv.org/abs/1509.01031}{{\tt
  1509.01031}}].

\bibitem{Marrani:2017ihg}
A.~Marrani, \emph{{Non-Linear Invariance of Black Hole Entropy}},
  \href{http://dx.doi.org/10.22323/1.314.0543}{\emph{PoS} {\bf EPS-HEP2017}
  (2017) 543}.

\bibitem{Marrani:2019zsn}
A.~Marrani, \emph{Non-linear symmetries in maxwell-einstein gravity: From
  freudenthal duality to pre-homogeneous vector spaces},  in \emph{Lie Theory
  and Its Applications in Physics} (V.~Dobrev, ed.), (Singapore), pp.~253--264,
  Springer Singapore, 2020.

\bibitem{Chattopadhyay:2021vog}
A.~Chattopadhyay and T.~Mandal, \emph{{Freudenthal duality of near-extremal
  black holes and Jackiw-Teitelboim gravity}},
  \href{http://dx.doi.org/10.1103/PhysRevD.105.046014}{\emph{Phys. Rev. D} {\bf
  105} (2022) 046014}, [\href{http://arxiv.org/abs/2110.05547}{{\tt
  2110.05547}}].

\bibitem{Chattopadhyay:2022ycb}
A.~Chattopadhyay, T.~Mandal and A.~Marrani, \emph{{Near-extremal Freudenthal
  duality}}, \href{http://dx.doi.org/10.1007/JHEP08(2023)014}{\emph{JHEP} {\bf
  08} (2023) 014}, [\href{http://arxiv.org/abs/2212.13500}{{\tt 2212.13500}}].

\bibitem{Astefanesei:2006dd}
D.~Astefanesei, K.~Goldstein, R.~P. Jena, A.~Sen and S.~P. Trivedi,
  \emph{{Rotating attractors}},
  \href{http://dx.doi.org/10.1088/1126-6708/2006/10/058}{\emph{JHEP} {\bf 10}
  (2006) 058}, [\href{http://arxiv.org/abs/hep-th/0606244}{{\tt
  hep-th/0606244}}].

\bibitem{Ferrara:2008hwa}
S.~Ferrara, K.~Hayakawa and A.~Marrani, \emph{{Lectures on Attractors and Black
  Holes}}, \href{http://dx.doi.org/10.1002/prop.200810569}{\emph{Fortsch.
  Phys.} {\bf 56} (2008) 993--1046},
  [\href{http://arxiv.org/abs/0805.2498}{{\tt 0805.2498}}].

\bibitem{TheFibonacciSequenceandtheGoldenRatio}
V.~P. Schielack, \emph{The fibonacci sequence and the golden ratio},
  \href{http://dx.doi.org/10.5951/MT.80.5.0357}{\emph{The Mathematics Teacher}
  {\bf 80} (1987) 357 -- 358}.

\bibitem{Kunduri:2007vf}
H.~K. Kunduri, J.~Lucietti and H.~S. Reall, \emph{{Near-horizon symmetries of
  extremal black holes}},
  \href{http://dx.doi.org/10.1088/0264-9381/24/16/012}{\emph{Class. Quant.
  Grav.} {\bf 24} (2007) 4169--4190},
  [\href{http://arxiv.org/abs/0705.4214}{{\tt 0705.4214}}].

\bibitem{Bardeen:1999px}
J.~M. Bardeen and G.~T. Horowitz, \emph{{The Extreme Kerr throat geometry: A
  Vacuum analog of AdS(2) x S**2}},
  \href{http://dx.doi.org/10.1103/PhysRevD.60.104030}{\emph{Phys. Rev. D} {\bf
  60} (1999) 104030}, [\href{http://arxiv.org/abs/hep-th/9905099}{{\tt
  hep-th/9905099}}].

\bibitem{GaussianParabolicandHyperbolicNumbers}
W.~Miller and R.~Boehning, \emph{Gaussian, parabolic, and hyperbolic numbers},
  \href{http://dx.doi.org/10.5951/MT.61.4.0377}{\emph{The Mathematics Teacher}
  {\bf 61} (1968) 377 -- 382}.

\bibitem{Kestelman}
H.~Kestelman, \emph{Automorphisms of the field of complex numbers},
  \href{http://dx.doi.org/https://doi.org/10.1112/plms/s2-53.1.1}{\emph{Proceedings
  of the London Mathematical Society} {\bf s2-53} (1951) 1--12},
  [\href{http://arxiv.org/abs/https://londmathsoc.onlinelibrary.wiley.com/doi/pdf/10.1112/plms/s2-53.1.1}{{\tt
  https://londmathsoc.onlinelibrary.wiley.com/doi/pdf/10.1112/plms/s2-53.1.1}}].

\bibitem{LopesCardoso:2007hen}
G.~Lopes~Cardoso, J.~M. Oberreuter and J.~Perz, \emph{{Entropy function for
  rotating extremal black holes in very special geometry}},
  \href{http://dx.doi.org/10.1088/1126-6708/2007/05/025}{\emph{JHEP} {\bf 05}
  (2007) 025}, [\href{http://arxiv.org/abs/hep-th/0701176}{{\tt
  hep-th/0701176}}].

\bibitem{Goldstein:2007km}
K.~Goldstein and R.~P. Jena, \emph{{One entropy function to rule them all...}},
  \href{http://dx.doi.org/10.1088/1126-6708/2007/11/049}{\emph{JHEP} {\bf 11}
  (2007) 049}, [\href{http://arxiv.org/abs/hep-th/0701221}{{\tt
  hep-th/0701221}}].

\end{thebibliography}\endgroup
{}

\end{document}